\documentclass[prc,showpacs,amsmath,amsfonts,eqsecnum,twocolumn]{revtex4}

\usepackage{graphicx}
\usepackage{pstcol}
\usepackage{bm}
\begin{document}
\hyphenation{Rijken}
\hyphenation{Nijmegen}
 
\title{
    ESC NN-Potentials in Momentum Space \\ 
    II. Meson-Pair Exchange Potentials }
\author{Th.A.\ Rijken}
\author{H.\ Polinder} 
\affiliation{Institute for Theoretical Physics Nijmegen, University of Nijmegen,
         Nijmegen, The Netherlands }
\author{ J. Nagata}                             
\altaffiliation{Present address: Kyushu International University, Fukuoka 805-8512, Japan}                     
\affiliation{Venture Business Laboratory, Hiroshima University, Kagamiyama 2-313, 
         Higashi-Hiroshima, Japan } 
\date{version of: \today}
 
\begin{abstract}

The partial wave projection of the Nijmegen soft-core potential model
for Meson-Pair-Exchange (MPE) for $NN$-scattering in momentum space is presented.
Here, nucleon-nucleon momentum space MPE-potentials are $NN$-interactions where
either one or both nucleons contains a meson-pair vertex.
Dynamically, the meson-pair vertices can be viewed as describing 
in an effective way (part of) the effects of heavy-meson
exchange and meson-nucleon resonances. {}From the point of view
of ``duality,'' these two kinds of contribution are roughly
equivalent. 
Part of the MPE-vertices can be found in the 
chiral-invariant phenomenological Lagrangians that have a 
basis in spontaneous broken chiral symmetry. 
It is shown that the MPE-interactions are a very important component of the
nuclear force, which indeed enables a very succesful description of 
the low and medium energy $NN$-data.
Here we present a precise fit to the $NN$-data with the extended-soft-core
(ESC) model containing OBE-, PS-PS-, and MPE-potentials. An excellent description 
of the $NN$-data for $T_{Lab} \leq 350$ MeV is presented and discussed.
Phase shifts are given and a $\chi^2_{p.d.p.} = 1.15$ is reached.
\end{abstract}

\maketitle
 
 
\section{Introduction}                                     
\label{sec:1}
In the previous paper \cite{RN00a}, henceforth refered to as paper I,  
the techniques for 
the momentum space treatment of the extended soft-core model,
hereafter referred to as the ESC-model, are described. 
This implies first the development of a representation
of the ESC-model suitable for the projection onto the Pauli-spinor
rotational invariant operators, and secondly the partial wave analysis.
This partial wave analysis is organized along similar lines as used 
for the soft-core OBE-models \cite{RKS91}.
In \cite{RN00a} the nucleon-nucleon partial wave contributions have 
been worked out in detail. These are the analogs of the configuration
space two-meson-exchange (TME) potentials given in e.g. \cite{RS96a}.
Here, the TME-potentials are defined to contain the planar- and crossed-box 
two-meson-exchange potentials.

In this second paper on soft-core two-meson-exchange potentials in 
momentum space, occasiously refered to as paper II in the following,
we derive the same representation as in paper I, but now for the
contributions to the nucleon-nucleon potentials when either one or
both nucleons contains a pair vertex, i.e. the MPE-potentials.
We give the partial wave potentials in the similar representation as used in paper I.
In Ref.~\cite{RS96b} the MPE-contributions to the configuration space
nucleon-nucleon potentials, i.e. when either one or
both nucleons contains a pair vertex, have been derived. 
The corresponding ``seagull'' diagrams are refered to as one-pair and two-pair diagrams. 
This in order to distinct these from the planar and crossed-box diagrams, 
which were given Ref.~\cite{RS96a}.   

The two types of two-meson-exchange potentials TME, see I, and MPE presented here 
are part of our program to extend
the Nijmegen soft-core one-boson-exchange potential~\cite{NRS78,MRS89,Rij85} to
arrive at a new extended soft-core nucleon-nucleon model, hereafter
referred to as the ESC potential~\cite{Rij93,RS96b,SR97,Rij99}.
 
In the introduction to Ref.~\cite{RS96b} a rather complete description
is given of the physical background behind the MPE-potentials,
and we refer the interested reader to that reference.

We apply the potentials derived in this work to fit the $NN$-data. 
In the TME-potentials we restrict ourselves to the ps-ps exchange. Or, 
phrased differently, we include only the Goldstone-boson sector. 
This because it gives the complete long-range contribution, OPEP+TPEP 
and the inclusion of $\eta$ etc. is necessary for (i) (approximate) 
chiral symmetry, and (ii) for completeness in the sense of $SU_f(3)$, which
allows an extension to hyperon-nucleon and hyperon-hyperon \cite{Rij99}.

\noindent In fact, this fit has been performed in the configuration space 
version. However, the results were checked numerically in momentum space, using 
the formulas of papers I and II.

This paper is organized as follows.
In section II and III, we give the essentials of the procedure followed in 
deriving the new momentum space representation.
In section IV the projection of the MPE on the Pauli-spinor invariants
is worked out for the adiabatic contributions.
In section V the same is done for the $1/M$-corrections: the non-adiabatic and 
the pseudo-vector-vertex terms.
In section VI the partial wave analysis is indicated. The procedure for the
partial wave projection is completely analogous to that of paper I, and can
be transcribed immediately comparing the invariant contributions 
$\Omega_j({\bf k}^2;t,u)$ for MPE to those for TME in I.
In section VII the results from a fit to the $NN$-data are shown and discussed.
Here, phase shifts are given for $T_{Lab} \leq 350$ MeV and the pair-couplings
are compared to the values expected from e.g. chiral lagrangians.

In Appendix A the pair-interaction Hamiltonians are listed.                    
In Appendix B the $\lambda$-representations for the MPE-denominators are given.
In Appendix C we give the integration dictionary for the 
gaussian integrals that occur in MPE but not in TME.
In Appendix D a derivation for the potentials due to the 'derivative scalar
pair' interaction, see the $g'_{(\pi\pi)_0}$-coupling in \ref{appAA.1a} 
is outlined. This for completeness, since although we do not employ this
kind of pair interaction, it occurs occurs often in the current literature.
In Appendix E the full $SU_f(3)$ contents of our pair interactions is shown.   
 
\begin{widetext}

\section{ Momentum Space Representation MPE-Potentials}        
\label{sec:2}
Here, we give an outline the essentials of the procedure to derive our
new momentum space representation for the MPE-potentilas. These procedures
have been described in I, to which we refer for details. Here, we focuss
on the peculiar features that occur in the application to the MPE-potentials.

The starting point is the basic convolutive integral 
\begin{eqnarray}
\widetilde{V}_{M,N} ({\bf k}) &=&  
 \int\!\!\int\!\frac{d^{3}k_{1}d^{3}k_{2}}{(2\pi)^{3}}   
 \delta({\bf k}-{\bf k}_1-{\bf k}_2)\
 \tilde{F}_{M}({\bf k}_{1}^{2},m_1)\tilde{G}_{N}({\bf k}_{2}^{2},m_2) 
 \nonumber\\
 &=& \int\!\!\frac{d^3\Delta}{(2\pi)^3} 
\tilde{F}_M(\mbox{\boldmath $\Delta$}^2,m_1)\ 
\tilde{G}_N(({\bf k}-\mbox{\boldmath $\Delta$})^2,m_2)\ , \nonumber\\
   && \label{eq:2.1}   
\end{eqnarray}
where $\tilde{F}_M({\bf k}^2)$ and $\tilde{G}_N({\bf k}^2)$ can be of the form
\begin{eqnarray}
 M=0:\ \ \tilde{F}_0({\bf k}^2) &=& \exp\left[-{\bf k}^2/\Lambda_1^2\right]\ \ , \ \
 M=2:\ \ \tilde{F}_2({\bf k}^2)  =  \frac{\exp\left[-{\bf k}^2/\Lambda_1^2\right]}        
 {{\bf k}^2+m_1^2}\ , \nonumber\\
 N=0:\ \ \tilde{G}_0({\bf k}^2) &=& \exp\left[-{\bf k}^2/\Lambda_2^2\right]\ \ , \ \
 N=2:\ \ \tilde{G}_2({\bf k}^2)  =  \frac{\exp\left[-{\bf k}^2/\Lambda_2^2\right]}        
 {{\bf k}^2+m_2^2}\ , \nonumber\\
   && \label{eq:2.2}   
\end{eqnarray}
i.e. $M,N=2$ is the modified Yukawa type and $M,N=0$ is the Gaussian type. Below, we 
give for the different cases the momentum space representation, similar to the 
one that has been developed in paper I:\\

\noindent (i) $M=N=2$: In paper I using twice the identity
\begin{equation}
 \frac{\exp\left[-{\bf k}^2/\Lambda^2\right]}{{\bf k}^2+m^2} =             
 e^{m^2/\Lambda^2} \int_1^\infty\!\!\frac{dt}{\Lambda^2} \exp\left[-\left(
 \frac{{\bf k}^2+m^2}{\Lambda^2}\right) t \right]
\label{eq:2.3} \end{equation}
the $\Delta$-integral has been carried out. After a redefinition of the variables
$t \rightarrow t/\Lambda_1^2$ and $u \rightarrow u/\Lambda_2^2$ the result in I is
\begin{eqnarray}
 \tilde{V}_{2,2}({\bf k})&=&(4\pi)^{-3/2} e^{m_1^2/\Lambda_1^2} e^{m_2^2/\Lambda_2^2}
 \int_{t_0}^\infty\!\!dt \int_{u_0}^\infty\!\! du \frac{\exp[-(m_1^2 t + m_2^2 u)]}
 {(t+u)^{3/2}}\cdot \nonumber\\
 && \times \exp\left[-\left(\frac{tu}{t+u}\right){\bf k}^2\right]\ \ (t_0=1/\Lambda_1^2, 
 u_0=1/\Lambda_2^2)\ .
\label{eq:2.4} \end{eqnarray}

\noindent (ii) $M=2, N=0$: Using the identity (\ref{eq:2.3}) once, and performing 
similar steps as in paper I, one easily derives that for this case
\begin{eqnarray}
 \tilde{V}_{2,0}({\bf k})&=&(4\pi)^{-3/2} e^{m_1^2/\Lambda_1^2} e^{m_2^2/\Lambda_2^2}
 \int_{t_0}^\infty\!\!dt \int_{u_0}^\infty\!\! du \frac{\exp[-(m_1^2 t + m_2^2 u)]}
 {(t+u)^{3/2}}\cdot \nonumber\\
 && \times \exp\left[-\left(\frac{tu}{t+u}\right){\bf k}^2\right]\cdot \delta(u-u_0)\ .  
\label{eq:2.5} \end{eqnarray}
Here, is defined $\delta(u-u_0) \equiv \lim_{\epsilon \downarrow 0}\delta(u-u_{0,\epsilon})$,
where $u_{0,\epsilon}=u_0-\epsilon$. This definition implies that in (\ref{eq:2.5}) the
$u$-integration can simply be performed by the substitution $u \rightarrow u_0$ in the 
integrand.

\noindent (iii) $M=0, N=2$: Similarly to the previous case, one has                 
\begin{eqnarray}
 \tilde{V}_{0,2}({\bf k})&=&(4\pi)^{-3/2} e^{m_1^2/\Lambda_1^2} e^{m_2^2/\Lambda_2^2}
 \int_{t_0}^\infty\!\!dt \int_{u_0}^\infty\!\! du \frac{\exp[-(m_1^2 t + m_2^2 u)]}
 {(t+u)^{3/2}}\cdot \nonumber\\
 && \times \exp\left[-\left(\frac{tu}{t+u}\right){\bf k}^2\right]\cdot \delta(t-t_0)\ .  
 \nonumber\\ && 
\label{eq:2.6} \end{eqnarray}
For the integrals $\tilde{V}_{M,N}$ of this section, and similar integrals below in 
this paper, we introduce the following convenient short-hand notation. We write
\begin{subequations}
\begin{eqnarray}
 \tilde{V}_{M,N}({\bf k})&=& \int_{t_0}^\infty\!\!dt \int_{u_0}^\infty\!\! du\ w(t,u)\cdot
 \left\{v_{M,N}(t,u)\ \exp\left[-\left(\frac{tu}{t+u}\right){\bf k}^2\right]\right\}\ ,  
\label{eq:2.7a} \end{eqnarray}
with common weight function $w_0(t,u)$ defined as
\begin{equation}
 w_0(t,u) \equiv (4\pi)^{-3/2} e^{m_1^2/\Lambda_1^2} e^{m_2^2/\Lambda_2^2}
 \frac{\exp[-(m_1^2 t + m_2^2 u)]}{(t+u)^{3/2}}\ .                
\label{eq:2.7b} \end{equation}
\end{subequations}

\noindent The form in which these basic integrals appear in MPE depends on two factors:
\begin{enumerate}
\item[(i)] The denominators $D(\omega_1,\omega_2)$. In the next section we will 
give a catalogue of these.
\item[(ii)] The operators $\tilde{O}({\bf k}_1,{\bf k}_2)$. Also these will be   
given in the next section.
\end{enumerate}

\end{widetext}




\section{ Meson-Pair Exchange Potentials}        
\label{sec:3}
In \cite{RS96b} the derivation of
the pair-exchange potentials both in momentum and in configuration space is given.       
In this reference the configuration space potentials are worked out fully.
The topic of this paper is to do the same for the momentum space description.
In particular, the partial wave analysis is performed leading to a 
representation which is very suitable for numerical evaluation.\\

 \begin{figure}   
 \resizebox{5.cm}{11.25cm}        
 {\includegraphics[200,200][400,650]{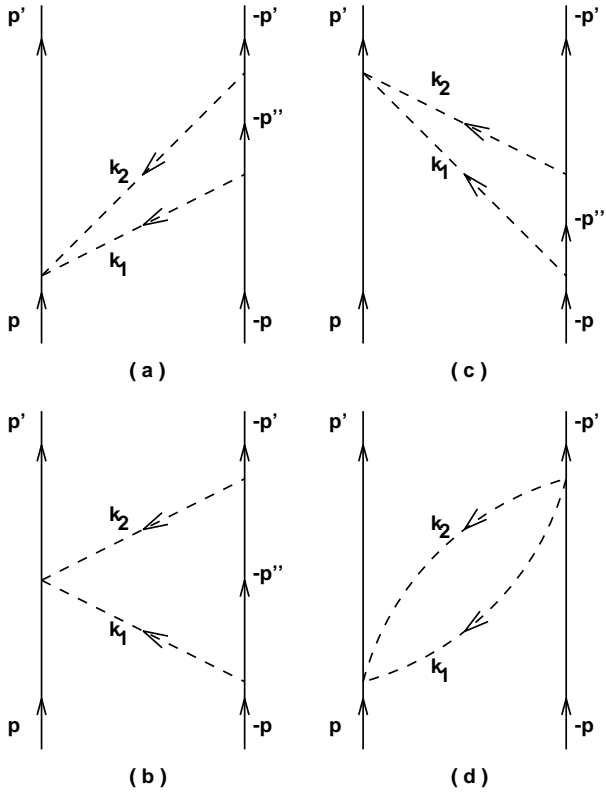}}
\caption{Time-ordered (a)-(c) one-pair and (d) two-pair diagrams. The dashed line with momentum $\textbf{k}_1$ refers to the pion and the dashed line with momentum $\textbf{k}_2$ refers to one of the other (vector, scalar, or pseudoscalar) mesons. To these we have to add the ``mirror'' diagrams, where for the one-pair diagrams the pair vertex occurs on the other nucleon line.}
\label{bwfig}
 \end{figure}

{}From \cite{RS96b} and equation (\ref{eq:3.1}) it follows that the momentum space 
MPE-potential can be represented in general in the form
\begin{eqnarray}
\widetilde{V}_{\alpha\beta}^{(n)} ({\bf k}) &=&  
 C^{(n)}(\alpha\beta) g^{(n)}(\alpha\beta)
 \int\!\!\int\!\frac{d^{3}k_{1}d^{3}k_{2}}{(2\pi)^{3}}   \nonumber\\
 && \times \delta({\bf k}-{\bf k}_1-{\bf k}_2)\
 \tilde{F}_{0}({\bf k}_{1}^{2})\tilde{G}_{0}({\bf k}_{2}^{2}) \nonumber\\
 && \times \sum_{p}\tilde{O}^{(n)}_{\alpha\beta,p}({\bf k}_{1},{\bf k}_2)\
          D_{p}^{(n)}(\omega_{1},\omega_{2}),
                                                            \nonumber\\
   && \label{eq:3.1}   
\end{eqnarray}
where the index $n$ distinguishes one-pair ($n=1$) and two-pair
($n=2$) meson-pair exchange, and ($\alpha\beta$) refers to the particular
meson pair that is being exchanged. 
The subscript $p=\{ad,na,pv\}$ distinguishes respectively
the adiabatic-, the non-adiabatic-, 
pseudovector vertex-, and off-shell-contributions. Here, the last three are 
the $1/M$-corrections to the MPE-potentials.\\
The product of the coupling constants in the cases $n=1,2$ is given by
\begin{eqnarray}
   g^{(1)}(\alpha\beta) &=& g_{(\alpha\beta)}
                             g_{N\!N\alpha}g_{N\!N\beta}, \nonumber\\
   g^{(2)}(\alpha\beta) &=& g^{2}_{(\alpha\beta)},
\label{eq:3.2} \end{eqnarray}
with appropriate powers of $m_{\pi}$, depending on the definition
of the Hamiltonians given in \cite{RS96b}, section II.

The momentum-dependent
operators $O^{(n)}_{\alpha\beta,p}$ are given in Tables~\ref{O1pair} and
\ref{O2pair}. For completeness, these Tables also contain the isospin
factors $C^{(n)}(\alpha\beta)$ as derived in Appendix B of \cite{RS96b}.
The momentum operators for $(\pi\pi)_{0}$ and $(\pi\pi)_{1}$ both
contain a term antisymmetric in ${\bf k}_{1}\leftrightarrow{\bf k}_{2}$, 
which only contributes in the nonadiabatic contribution, see \cite{RS96b},
section 4. In the adiabatic potential, as explained in \cite{RS96b}, 
they drop out when we integrate over ${\bf k}_{1}$ and ${\bf k}_{2}$.
 
The energy denominators $D^{(n)}_{p}$ are also discussed in detail in 
\cite{RS96b}), section II, in terms of the time-ordered processes involved in 
one- and two-pair exchange. These denominators depend on the energies 
of the exchanged mesons, i.e. $\omega_1$ and $\omega_2$. Another sourse of
$\omega_{1,2}$-dependence comes from vertices with derivatives, and the
non-adiabatic expansion terms. It appears from \cite{RS96b} that in 
general one can write
\begin{eqnarray}
 D_{\{p\}}^{(n)}(\omega_{1},\omega_{2}) &=& \sum_{p_1,p_2,p_3} 
 c^{(n)}_{p_1,p_2,p_3}\ D_{\{p_1,p_2,p_3\}}\ , 
\label{eq:3.3} \end{eqnarray}
where in terms of the integer powers $p_i\ (i=1,2,3)$ the denominators 
can be written 
\begin{equation}
 D_{\{p_1,p_2,p_3\}} = \frac{1}{\omega_1^{p_1}} \frac{1}{\omega_2^{p_2}}
 \frac{1}{(\omega_1+\omega_2)^{p_3}}\ .
\label{eq:3.4} \end{equation}
The energy denominators $D^{(n)}_p$ are listed in Tables~\ref{D1pair} and
\ref{O2pair}. 

The evaluation of the momentum integrations can now readily be
performed using the methods given in~\cite{Rij91,RS96b}.
There it was shown that the full separation of the ${\bf k}_{1}$
and ${\bf k}_{2}$ dependence can be
achieved in all cases using the $\lambda$-integral representation, 
first introduced in  \cite{Rij91}. In Appendix \ref{app:A} the 
occurring $\lambda$-integrals are listed. {}From the listing in \ref{app:A}
one readily sees that for the derivation of the representation similar to
that one in Eqs.~(\ref{eq:2.4})-(\ref{eq:2.6})
we need to start out from the generalization of (\ref{eq:2.1}):
\begin{eqnarray}
&&\widetilde{V}_{M,N} ({\bf k},\lambda) = \frac{2}{\pi}\int_0^\infty\!\!
 d\lambda f_{M,N}(\lambda) 
 \int\!\!\frac{d^3\Delta}{(2\pi)^3} 
\cdot \nonumber\\ && \times
\tilde{F}_M(\mbox{\boldmath $\Delta$}^2,\sqrt{m_1^2+\lambda^2})\ 
\tilde{G}_N(({\bf k}-\mbox{\boldmath $\Delta$})^2,\sqrt{m_2^2+\lambda^2})\ . 
\label{eq:3.5} \end{eqnarray}
In paper I it has been shown that all the occurring $\lambda$-integrals can
be performed analytically. The result for all cases can be written as

\begin{widetext}
 
\begin{eqnarray}
\widetilde{V}_{p_1,p_2,p_3} ({\bf k}) &=&  \int\!\!\frac{d^3\Delta}{(2\pi)^3} 
\tilde{F}_0(\mbox{\boldmath $\Delta$}^2,m_1)\ 
\tilde{G}_0(({\bf k}-\mbox{\boldmath $\Delta$})^2,m_2)\ 
          D_{\{p_1,p_2,p_3\}}(\omega_{1},\omega_{2}) \nonumber\\
&=& \int_{t_0}^\infty\!\!dt \int_{u_0}^\infty\!\! du\ w_0(t,u)\cdot
\left\{d_{\{p_1,p_2,p_3\}}(t,u)\ 
\exp\left[-\left(\frac{tu}{t+u}\right){\bf k}^2\right]\right\}\ .  
\label{eq:3.6} \end{eqnarray}
\end{widetext}


All functions $d_{\{p_1,p_2,p_3\}}(t,u)$ that occur in this work are given 
in Table~\ref{dtus}. As noted in section~\ref{sec:2} we will use only representations
with $M=N=2$, so that no $\delta(t-t_0)$ or $\delta(u-u_0)$ occurs.

\begin{widetext}

\section{ Projection MPE on Spinor Invariants I\protect\\
    Adiabatic Contributions }
\label{sec:4}
The MPE-contributions from the adiabatic terms, the non-adiabatic- and
pseudovector vertex corrections are the central-, spin-spin-, tensor-, 
and spin-orbit momentum space analogs of those given in Reference~\cite{RS96b}
in configuration space.
{}From (\ref{eq:3.1}), Tables~\ref{O1pair}-\ref{O2pair}, and Table~\ref{D1pair}
it is readily verified that the projection onto the potentials $V_j$ , similarly
to the paper I, can be written as
\begin{eqnarray}
    \widetilde{V}^{(n)}_{\rm pair}(\alpha\beta) &=&
 \int_{t_0}^\infty\!\!dt \int_{u_0}^\infty\!\! du\ w_0(t,u)\ \left\{
\exp\left[-\left(\frac{tu}{t+u}\right){\bf k}^2\right]\   
 \Omega^{(n)}_j({\bf k}^2;t,u)\right\}(\alpha\beta)\ .
\label{eq:4.1} \end{eqnarray}
The functions $\Omega^{(n)}_j$ are worked out in the subsections below.
Like in I, we also introduce for convenience the expansion in ${\bf k}^2$:
\begin{equation}
\Omega_j^{(ad,na,pv)}({\bf k}^2;t,u) = C^{(n)}(\alpha\beta) g^{(n)}(\alpha\beta)\cdot
 \sum_{k=0}^K\ \Upsilon_{j,k}^{(ad,na,pv)}(t,u)\ \left({\bf k}^2\right)^k\ .
\label{eq:4.1b} \end{equation}
Below in this section we give the results for the adiabatic contributions.  
The coefficients $\Upsilon_{j,k}^{ad}$ are tabulated in Tables~\ref{table1}-\ref{table5}.

\subsection{ $J^{PC}=0^{++}$: Adiabatic $(\pi\pi)_0$-Exchange potentials}   
The 1-pair and 2-pair contributions are
\begin{subequations}
\begin{eqnarray}
\Omega^{(1)}_1({\bf k}^2;t,u) &=& 6                    
\left(\frac{g_{(\pi\pi)_0}}{m_\pi}\right)
 \left(\frac{f^2_{NN\pi}}{m_\pi^2}\right)\cdot d_{\{2,2,0\}}(t,u)
\left\{+\frac{3}{2}-\left(\frac{tu}{t+u}\right){\bf k}^2\right\}\cdot\frac{1}{t+u}\ ,
\label{eq:4.2a}\\
\Omega^{(2)}_1({\bf k}^2;t,u) &=& -3
\left(\frac{g_{(\pi\pi)_0}}{m_\pi^2}\right)^2\cdot d_{\{1,1,1\}}(t,u)
\label{eq:4.2b}
\end{eqnarray}
\end{subequations}

\subsection{ $J^{PC}= 1^{--}$: Adiabatic $(\pi\pi)_1$-Exchange potentials}   
\noindent (i)\ 1-pair exchange:
\begin{subequations}
\begin{eqnarray}
\Omega^{(1)}_1({\bf k}^2;t,u) &=& -4 
(\mbox{\boldmath $\tau$}_1\cdot\mbox{\boldmath $\tau$}_2)
\left(\frac{g_{(\pi\pi)_1}}{m_\pi^2}\right)
 \left(\frac{f^2_{NN\pi}}{m_\pi^2}\right)\cdot d_{1,1,1}(t,u)
\left\{-\frac{3}{2}+\left(\frac{tu}{t+u}\right){\bf k}^2\right\}\cdot\frac{1}{t+u}\ ,
\label{eq:4.3a}\\
\Omega^{(1)}_2({\bf k}^2;t,u) &=& -2 
(\mbox{\boldmath $\tau$}_1\cdot\mbox{\boldmath $\tau$}_2)
\left(\frac{g_{(\pi\pi)_1}}{m_\pi^2}\right) \left(\frac{f^2_{NN\pi}}{m_\pi^2}\right)
\frac{(1+\kappa_1)}{M}\cdot d_{2,2,0}(t,u)\cdot
+\frac{1}{3}{\bf k}^2\cdot\frac{1}{t+u}\ ,
\label{eq:4.3b}\\
\Omega^{(1)}_3({\bf k}^2;t,u) &=& -2 
(\mbox{\boldmath $\tau$}_1\cdot\mbox{\boldmath $\tau$}_2)
\left(\frac{g_{(\pi\pi)_1}}{m_\pi^2}\right) \left(\frac{f^2_{NN\pi}}{m_\pi^2}\right)
\frac{(1+\kappa_1)}{M}\cdot d_{2,2,0}(t,u)\cdot
-\frac{1}{2}\cdot\frac{1}{t+u}\ ,
\label{eq:4.3c}\\
\Omega^{(1)}_4({\bf k}^2;t,u) &=& -2 
(\mbox{\boldmath $\tau$}_1\cdot\mbox{\boldmath $\tau$}_2)
\left(\frac{g_{(\pi\pi)_1}}{m_\pi^2}\right) \left(\frac{f^2_{NN\pi}}{m_\pi^2}\right)
\frac{1}{M}\cdot d_{2,2,0}(t,u) \cdot\frac{1}{t+u}\ ,
\label{eq:4.3d}
\end{eqnarray}

\noindent (ii)\ 2-pair exchange:
\begin{eqnarray}
\Omega^{(2)}_1({\bf k}^2;t,u) &=& -\frac{1}{2} 
(\mbox{\boldmath $\tau$}_1\cdot\mbox{\boldmath $\tau$}_2)
\left(\frac{g_{(\pi\pi)_1}}{m_\pi^2}\right)^2\cdot 
\left[d_{1,0,0}+d_{0,1,0}-4d_{0,0,1}\right](t,u)\ .
\label{eq:4.3e}
\end{eqnarray}
\end{subequations}

\subsection{ $J^{PC}= 1^{++}$: Adiabatic $(\pi\rho)_1$-Exchange potentials}   
\noindent (i)\ 1-pair exchange:
\begin{subequations}
\begin{eqnarray}
\Omega^{(1)}_2({\bf k}^2;t,u) &=& \frac{2}{M}\
(\mbox{\boldmath $\tau$}_1\cdot\mbox{\boldmath $\tau$}_2)
\left(\frac{g_{(\pi\rho)_1}}{m_\pi}\right)
 \left(\frac{f_{NN\pi}g_{NN\rho}}{m_\pi}\right)\cdot d_{2,2,0}(t,u)
\cdot \nonumber\\ && \times
\left\{\left[\frac{1}{2}+\frac{1}{3}\left(\frac{u^2}{t+u}\right){\bf k}^2\right] 
+\frac{1}{2}(1+\kappa_\rho)\left[2-\frac{4}{3}\frac{tu}{t+u}{\bf k}^2\right]\right\}
\cdot\frac{1}{t+u}\ ,
\label{eq:4.4a}\\
\Omega^{(1)}_3({\bf k}^2;t,u) &=& \frac{2}{M}\
(\mbox{\boldmath $\tau$}_1\cdot\mbox{\boldmath $\tau$}_2)
\left(\frac{g_{(\pi\rho)_1}}{m_\pi}\right)
 \left(\frac{f_{NN\pi}g_{NN\rho}}{m_\pi}\right)\cdot d_{2,2,0}(t,u)
\cdot \nonumber\\ && \times
\left\{\frac{u^2}{t+u}+\frac{1}{2}(1+\kappa_\rho)\frac{2tu}{t+u}\right\}
\cdot\frac{1}{t+u}\ ,
\label{eq:4.4b}
\end{eqnarray}

\noindent (ii)\ 2-pair exchange:
\begin{eqnarray}
\Omega^{(2)}_2({\bf k}^2;t,u) &=& - 
(\mbox{\boldmath $\tau$}_1\cdot\mbox{\boldmath $\tau$}_2)
\left(\frac{g_{(\pi\rho)_1}}{m_\pi^2}\right)^2\cdot d_{1,1,1}(t,u)
\label{eq:4.4c}
\end{eqnarray}
\end{subequations}

\subsection{ $J^{PC}= 1^{++}$: Adiabatic $(\pi\sigma)$-Exchange potentials}   
\noindent (i)\ 1-pair exchange:
\begin{subequations}
\begin{eqnarray}
\Omega^{(1)}_2({\bf k}^2;t,u) &=& +
(\mbox{\boldmath $\tau$}_1\cdot\mbox{\boldmath $\tau$}_2)
\left(\frac{g_{(\pi\sigma)}}{m_\pi^2}\right)
 \left(\frac{f_{NN\pi}g_{NN\sigma}}{m_\pi}\right)\cdot d_{2,2,0}(t,u)
\cdot \nonumber\\ && \times
\left[-2+\frac{2}{3}\left(\frac{tu-u^2}{t+u}\right){\bf k}^2\right] 
\cdot\frac{1}{t+u}\ ,
\label{eq:4.5a}\\
\Omega^{(1)}_3({\bf k}^2;t,u) &=& +2
(\mbox{\boldmath $\tau$}_1\cdot\mbox{\boldmath $\tau$}_2)
\left(\frac{g_{(\pi\sigma)}}{m_\pi^2}\right)
 \left(\frac{f_{NN\pi}g_{NN\sigma}}{m_\pi}\right)\cdot d_{2,2,0}(t,u)
\cdot \nonumber\\ && \times
\left(\frac{tu-u^2}{t+u}\right) 
\cdot\frac{1}{t+u}\ ,
\label{eq:4.5b}
\end{eqnarray}

\noindent (ii)\ 2-pair exchange:
\begin{eqnarray}
\Omega^{(2)}_2({\bf k}^2;t,u) &=& -\frac{1}{2}
(\mbox{\boldmath $\tau$}_1\cdot\mbox{\boldmath $\tau$}_2)
\left(\frac{g_{(\pi\sigma)_1}}{m_\pi^2}\right)^2\cdot d_{1,1,1}(t,u)
\cdot \nonumber\\ && \times
 \left\{\frac{2}{t+u}+\frac{1}{3}\left(\frac{t-u}{t+u}\right)^2{\bf k}^2\right\}\ ,
\label{eq:4.5c}\\
\Omega^{(2)}_3({\bf k}^2;t,u) &=& -\frac{1}{2}
(\mbox{\boldmath $\tau$}_1\cdot\mbox{\boldmath $\tau$}_2)
\left(\frac{g_{(\pi\sigma)_1}}{m_\pi^2}\right)^2\cdot d_{1,1,1}(t,u)
 \left(\frac{t-u}{t+u}\right)^2\ ,
\label{eq:4.5d}
\end{eqnarray}
\end{subequations}

\subsection{ $J^{PC}= 1^{+-}$: Adiabatic $(\pi\omega)$-Exchange potentials}   
\noindent (i)\ 1-pair exchange:
\begin{subequations}
\begin{eqnarray}
\Omega^{(1)}_2({\bf k}^2;t,u) &=& 
(\mbox{\boldmath $\tau$}_1\cdot\mbox{\boldmath $\tau$}_2)
\left(\frac{g_{(\pi\omega)}}{m_\pi}\right)
 \left(\frac{f_{NN\pi}g_{NN\omega}}{m_\pi}\right)\cdot d_{2,2,0}(t,u)
\cdot \nonumber\\ && \times
\left[\frac{2}{3}\left(\frac{tu+u^2}{t+u}\right){\bf k}^2\right] 
\cdot\frac{1}{t+u}\ ,
\label{eq:4.6a}\\
\Omega^{(1)}_3({\bf k}^2;t,u) &=& +2
(\mbox{\boldmath $\tau$}_1\cdot\mbox{\boldmath $\tau$}_2)
\left(\frac{g_{(\pi\omega)}}{m_\pi}\right)
 \left(\frac{f_{NN\pi}g_{NN\omega}}{m_\pi}\right)\cdot d_{2,2,0}(t,u)
\cdot \nonumber\\ && \times
\left(\frac{tu+u^2}{t+u}\right)\cdot\frac{1}{t+u}\ ,
\label{eq:4.6b}
\end{eqnarray}

\noindent (ii)\ 2-pair exchange:
\begin{eqnarray}
\Omega^{(2)}_2({\bf k}^2;t,u) &=& -\frac{1}{2}
(\mbox{\boldmath $\tau$}_1\cdot\mbox{\boldmath $\tau$}_2)
\left(\frac{g_{(\pi\omega)_1}}{m_\pi^2}\right)^2\cdot                   
\nonumber\\ && \times
\left\{ d_{1,0,0}+d_{0,1,0}-\frac{1}{3}{\bf k}^2 d_{1,1,1}\right\}(t,u)\ ,
\label{eq:4.6c}\\
\Omega^{(2)}_3({\bf k}^2;t,u) &=& +\frac{1}{2}
(\mbox{\boldmath $\tau$}_1\cdot\mbox{\boldmath $\tau$}_2)
\left(\frac{g_{(\pi\omega)_1}}{m_\pi^2}\right)^2\cdot d_{1,1,1}(t,u)\ .
\label{eq:4.6d}
\end{eqnarray}
\end{subequations}

\subsection{ $J^{PC}= 1^{++}$: Adiabatic $(\pi P)$-Exchange potentials}   
The treatment of the Pomeron has been explained in \cite{RS96a}. This implies the use 
of $\widetilde{G}_0/M_N^2$ in section~\ref{sec:2}. Furthermore, w.r.t. $\sigma$-exchange
there is a $(-)$-sign for P-exchange. Therefore, comparing to (\ref{eq:4.5a}-\ref{eq:4.5d})
we obtain the following potentials:

\noindent (i)\ 1-pair exchange:
\begin{subequations}
\begin{eqnarray}
\Omega^{(1)}_2({\bf k}^2;t,u) &=& -
(\mbox{\boldmath $\tau$}_1\cdot\mbox{\boldmath $\tau$}_2)
\left(\frac{g_{(\pi P)}}{m_\pi^2}\right)
 \left(\frac{f_{NN\pi}g_{NNP}}{m_\pi}\right)\cdot 
\frac{1}{M_N^2}\cdot d_{2,0,0}(t,u)\cdot \nonumber\\ && \times
\left[-2+\frac{2}{3}\left(\frac{tu-u^2}{t+u}\right){\bf k}^2\right] 
\cdot\frac{1}{t+u}\cdot \delta(u-u_0)\ ,
\label{eq:4.7a}\\
\Omega^{(1)}_3({\bf k}^2;t,u) &=& -2
(\mbox{\boldmath $\tau$}_1\cdot\mbox{\boldmath $\tau$}_2)
\left(\frac{g_{(\pi P)}}{m_\pi^2}\right)
 \left(\frac{f_{NN\pi}g_{NNP}}{m_\pi}\right)\cdot 
\frac{1}{M_N^2}\cdot d_{2,0,0}(t,u)\cdot \nonumber\\ && \times
\left(\frac{tu-u^2}{t+u}\right) 
\cdot\frac{1}{t+u}\cdot \delta(u-u_0)\ .
\label{eq:4.7b}
\end{eqnarray}

\noindent (ii)\ 2-pair exchange:
\begin{eqnarray}
\Omega^{(2)}_2({\bf k}^2;t,u) &=& +\frac{1}{2}
(\mbox{\boldmath $\tau$}_1\cdot\mbox{\boldmath $\tau$}_2)
\left(\frac{g_{(\pi P)_1}}{m_\pi}\right)^2
\cdot\frac{1}{M_N^2}\cdot d_{1,1,1}(t,u)\cdot \nonumber\\ && \times
 \left\{\frac{2}{t+u}+\frac{1}{3}\left(\frac{t-u}{t+u}\right)^2{\bf k}^2\right\}
\cdot\delta(u-u_0)\ ,
\label{eq:4.7c}\\
\Omega^{(2)}_3({\bf k}^2;t,u) &=& +\frac{1}{2}
(\mbox{\boldmath $\tau$}_1\cdot\mbox{\boldmath $\tau$}_2)
\left(\frac{g_{(\pi P)_1}}{m_\pi^2}\right)^2
\cdot\frac{1}{M_N^2}\cdot d_{1,1,1}(t,u) 
\left(\frac{t-u}{t+u}\right)^2\cdot\delta(u-u_0)\ . 
\label{eq:4.7d}
\end{eqnarray}
\end{subequations}
Notice that in (\ref{eq:4.7a})-(\ref{eq:4.7d}) $u_0 = 1/4m_P^2$.

\subsection{ $J^{PC}= 0^{++}$: 
 Adiabatic 'derivative' $(\pi\pi)_0$-Exchange potentials}   
 The derivative pair-potentials in coordinate space have been derived in 
\cite{Rij00} in detail. A summary of this is given in appendix \ref{app:E}.
A short derivation of the p-space potentials is also can be found there.

\noindent (i)\ 1-pair exchange:
\begin{subequations}
\begin{eqnarray}
\Omega^{(1)}_1({\bf k}^2;t,u) &=& -12
 \left(\frac{g'_{(\pi\pi)_0}}{m_\pi^3}\right)
 \left(\frac{f_{NN\pi}}{m_\pi}\right)^2\cdot 
 d_{2,2,0}(t,u)\cdot \nonumber\\ && \times
\left[\frac{15}{4}+\frac{1}{2}\frac{t^2-8tu+u^2}{t+u}{\bf k}^2
 + \frac{t^2u^2}{(t+u)^2}{\bf k}^4\right] 
\cdot\frac{1}{(t+u)^2}\ , \label{eq:4.8a}
\end{eqnarray}

\noindent (ii)\ 2-pair exchange:
\begin{eqnarray}
\Omega^{(2)}_1({\bf k}^2;t,u) &=& -6
 \left(\frac{g'_{(\pi\pi)_0}}{m_\pi^3}\right)^2\cdot
\left\{\left[\frac{15}{4}+\frac{t^2-3tu+u^2}{t+u}{\bf k}^2+
\frac{t^2u^2}{(t+u)^2}{\bf k}^4 \right]\frac{d_{1,1,1}(t,u)}{(t+u)^2} 
\right.\nonumber\\ && \left. 
 +\frac{1}{2}\left[\frac{3}{2}(m_1^2+m_2^2)+\frac{m_1^2 t + 
 m_2^2 u}{t+u}{\bf k}^2 +m_1^2m_2^2(t+u)\right] \frac{d_{1,1,1}(t,u)}{t+u}
\right.\nonumber\\ && \left.
 +\left[\frac{3}{2}-\frac{tu}{t+u}{\bf k}^2\right]\frac{d_{0,0,1}(t,u)}{t+u}\right\}\ .
\label{eq:4.8b}
\end{eqnarray}
\end{subequations}

\subsection{ $J^{PC}= 0^{++}$: 
 Adiabatic $(\sigma\sigma)$-Exchange potentials}   
\begin{subequations}
\noindent (i)\ 1-pair exchange:
\begin{eqnarray}
\Omega^{(1)}_1({\bf k}^2;t,u) &=& 2
 \left(\frac{g_{(\sigma\sigma)}}{m_\pi}\right)
 \cdot g_{NN\sigma}^2\cdot 
 d_{2,2,0}(t,u)\ , \label{eq:4.9a}
\end{eqnarray}

\noindent (ii)\ 2-pair exchange:
\begin{eqnarray}
\Omega^{(2)}_1({\bf k}^2;t,u) &=& -
 \left(\frac{g_{(\sigma\sigma)}}{m_\pi}\right)^2
 d_{1,1,1}(t,u)\ , \label{eq:4.9b}
\end{eqnarray}
\end{subequations}

\section{ Projection MPE on Spinor Invariants II \protect\\
    $1/M$ Corrections } 
\label{sec:5}
The non-adiabatic- and pseudovector vertex-corrections have been given in 
\cite{RS96b}, section IV. Similar to Eq.~(\ref{eq:4.1}) we write these 
contributions in the form
\begin{eqnarray}
    \widetilde{V}^{(na,pv)}_{\rm pair}(\alpha\beta) &=&
 \int_{t_0}^\infty\!\!dt \int_{u_0}^\infty\!\! du\ w_0(t,u)\
 \left\{d^{(n)}_{\{p_1,p_2,p_3\}}(t,u)\ 
\exp\left[-\left(\frac{tu}{t+u}\right){\bf k}^2\right]\   
 \Omega^{(n)}_j({\bf k}^2;t,u)\right\}\ .
\label{eq:5.1} \end{eqnarray}

\subsection{ Non-adiabatic Corrections}                           
\label{sec:5a}
{}From Eqs.~(4.5)-4.8) of \cite{RS96b} one readily obtains the momentum space  
equivalents using the replacements:
\[
 \int\int \frac{d^3k_1 d^3 k_2}{(2\pi)^6}\ e^{i({\bf k}_1+{\bf k}_2)}
 \rightarrow 
 \int\int \frac{d^3k_1 d^3 k_2}{(2\pi)^3}\ \delta({\bf k}-{\bf k}_1-{\bf k}_2)
\]
Then, by comparison one can easily read off the diverse quantities 
$\tilde{O}^{(na)}_{\alpha\beta,p}$ and $D^{(na)}_p(\omega_1,\omega_2)$ that occur 
in Eq. (\ref{eq:3.1}) for the non-adiabatic potentials. The projections onto 
the $\Omega^{(na)}_j$ in Eq.~(\ref{eq:5.1}) yield

\noindent (i)\ $(\pi\pi)_0$:
\begin{subequations}
\begin{eqnarray}
\Omega_1^{(na)}({\bf k}^2;t,u) &=& -\frac{g_{(\pi\pi)_0}}{m_\pi}
\left(\frac{f_{NN\pi}}{m_\pi}\right)^2\frac{3}{M}\cdot d_{\{na\}}(t,u)
\left\{\frac{15}{4}+
 \right.\nonumber\\ && \left. +
\frac{1}{2}\left(\frac{t^2-8ut+u^2}{t+u}\right) {\bf k}^2
+\left(\frac{t^2u^2}{(t+u)^2}\right){\bf k}^4\right\}\cdot\frac{1}{(t+u)^2}\ ,
\label{eq:5.2a}\\
\Omega_4^{(na)}({\bf k}^2;t,u) &=& -\frac{g_{(\pi\pi)_0}}{m_\pi}
\left(\frac{f_{NN\pi}}{m_\pi}\right)^2\frac{3}{M}\cdot d_{\{na\}}(t,u)
\cdot\frac{1}{t+u}\ .
\label{eq:5.2b} \end{eqnarray}
\end{subequations}

\noindent (ii)\ $(\pi\pi)_1$:
\begin{subequations}
\begin{eqnarray}
\Omega_1^{(na)}({\bf k}^2;t,u) &=& -2
(\mbox{\boldmath $\tau$}_1\cdot\mbox{\boldmath $\tau$}_2)
\frac{g_{(\pi\pi)_1}}{m_\pi}
\left(\frac{f_{NN\pi}}{m_\pi}\right)^2\frac{1}{M}\cdot d_{\{2,2,0\}}(t,u)
\left\{\frac{15}{4}+
 \right.\nonumber\\ && \left. +
\frac{1}{2}\left(\frac{t^2-8ut+u^2}{t+u}\right) {\bf k}^2
+\left(\frac{t^2u^2}{(t+u)^2}\right){\bf k}^4\right\}\cdot\frac{1}{(t+u)^2}\ ,
\label{eq:5.3a}\\
\Omega_4^{(na)}({\bf k}^2;t,u) &=& -2
(\mbox{\boldmath $\tau$}_1\cdot\mbox{\boldmath $\tau$}_2)
\frac{g_{(\pi\pi)_1}}{m_\pi}
\left(\frac{f_{NN\pi}}{m_\pi}\right)^2\frac{1}{M}\cdot d_{\{2,2,0\}}(t,u)
\cdot\frac{1}{t+u}\ .
\label{eq:5.3b} \end{eqnarray}
\end{subequations}

\noindent (iii)\ $(\sigma\sigma)$:
\begin{eqnarray}
\Omega_1^{(na)}({\bf k}^2;t,u) &=& 
\frac{g_{(\sigma\sigma)}}{m_\pi}
\frac{g_{NN\sigma}^2}{M}\cdot d_{\{na\}}(t,u)
\left\{-\frac{3}{2}+\left(\frac{tu}{t+u}\right) {\bf k}^2
\right\}\cdot\frac{1}{t+u}\ .
\label{eq:5.4} \end{eqnarray}

\noindent (iv)\ $(\pi\sigma)$:
\begin{subequations}
\begin{eqnarray}
\Omega_2^{(na)}({\bf k}^2;t,u) &=& +
(\mbox{\boldmath $\tau$}_1\cdot\mbox{\boldmath $\tau$}_2)
\frac{g_{(\pi\sigma)_1}}{m_\pi^2}
\frac{f_{NN\pi}}{m_\pi}\frac{g_{NN\sigma}}{M}\cdot d_{\{na\}}(t,u)
 \cdot \nonumber\\ && \times
\left\{\frac{5}{2}+\frac{1}{6}\frac{t^2-13tu+6u^2}{t+u}{\bf k}^2
 +\frac{1}{3}\frac{tu^2(t-u)}{(t+u)^2}{\bf k}^4\right\}\cdot\frac{1}{(t+u)^2}\ ,
\label{eq:5.5a}\\
\Omega_3^{(na)}({\bf k}^2;t,u) &=& +
(\mbox{\boldmath $\tau$}_1\cdot\mbox{\boldmath $\tau$}_2)
\frac{g_{(\pi\sigma)_1}}{m_\pi^2}
\frac{f_{NN\pi}}{m_\pi}\frac{g_{NN\sigma}}{M}\cdot d_{\{na\}}(t,u)
 \cdot \nonumber\\ && \times
\left\{\frac{1}{2}\frac{t^2-7tu+6u^2}{t+u}             
 +\frac{tu^2(t-u)}{(t+u)^2}{\bf k}^2\right\}\cdot\frac{1}{(t+u)^2}\ .
\label{eq:5.5b} \end{eqnarray}
\end{subequations}

\noindent (v)\ $(\pi\pi)_0$('derivative'):
\begin{subequations}
\begin{eqnarray}
&& \Omega_1^{(na)}({\bf k}^2;t,u) = -12
\left(\frac{g'_{(\pi\pi)_0}}{m_\pi^3}\right)
\left(\frac{f_{NN\pi}}{m_\pi}\right)^2\frac{1}{2M}\cdot 
 \left\{\vphantom{\frac{A}{A}}\right. \nonumber\\ & & +     
\left[\frac{15}{4}+\frac{1}{2}\left(\frac{t^2-8tu+u^2}{t+u}\right){\bf k}^2
      +\frac{t^2u^2}{(t+u)^2}{\bf k}^4\right]\frac{d_{1,1,1}(t,u)}{(t+u)^2}\ ,
 \nonumber\\ & & -\left.
\left[\frac{105}{8}+\frac{15}{4}\left(\frac{t^2-5tu+u^2}{t+u}\right){\bf k}^2
      -\frac{3}{2}tu\left(\frac{t^2-5tu+u^2}{(t+u)^2}\right){\bf k}^4
      -\frac{t^3u^3}{(t+u)^3}{\bf k}^6\right]\frac{d_{na}(t,u)}{(t+u)^3}
\right\}\ , \label{eq:5.5c}\\
 && \Omega_4^{(na)}({\bf k}^2;t,u) = - 12
\left(\frac{g'_{(\pi\pi)_0}}{m_\pi^3}\right)
\left(\frac{f_{NN\pi}}{m_\pi}\right)^2\frac{1}{2M}\cdot 
\left\{\frac{d_{1,1,1}(t,u)}{t+u}+  
\left[-5+2\frac{tu}{t+u}{\bf k}^2\right]
\frac{d_{na}(t,u)}{(t+u)^2} \right\}\ .
\label{eq:5.5d} \end{eqnarray}
\end{subequations}
 
Here, $d_{\{na\}}(t,u)$ is defined in (\ref{appA.3}).

\subsection{ Pseudovector-vertex Corrections}                      
\label{sec:5b}
{}From Eqs.~(4.9)-4.11) of \cite{RS96b} likewise as in the case of the  
non-adiabatic corrections one obtains for the pseudovector-vertex corrections:    

\noindent (i)\ $(\pi\pi)_0$:
\begin{subequations}
\begin{eqnarray}
\Omega_1^{(pv)}({\bf k}^2;t,u) &=& +\frac{g_{(\pi\pi)_0}}{m_\pi}
\left(\frac{f_{NN\pi}}{m_\pi}\right)^2\frac{3}{M}\cdot d_{\{1,1,1\}}(t,u)
\left\{ 3 +\left(\frac{t^2+u^2}{t+u}\right) {\bf k}^2
\right\}\cdot\frac{1}{t+u}\ ,
\label{eq:5.6a}\\
\Omega_4^{(pv)}({\bf k}^2;t,u) &=& +2\frac{g_{(\pi\pi)_0}}{m_\pi}
\left(\frac{f_{NN\pi}}{m_\pi}\right)^2\frac{3}{M}\cdot d_{\{1,1,1\}}(t,u)\ .
\label{eq:5.6b} \end{eqnarray}
\end{subequations}

\noindent (ii)\ $(\pi\pi)_1$:
\begin{subequations}
\begin{eqnarray}
\Omega_1^{(pv)}({\bf k}^2;t,u) &=& + 
(\mbox{\boldmath $\tau$}_1\cdot\mbox{\boldmath $\tau$}_2)
\frac{g_{(\pi\pi)_1}}{m_\pi}
\left(\frac{f_{NN\pi}}{m_\pi}\right)^2\frac{1}{M}\cdot 
\left[ \left( \frac{3}{2}+\frac{u^2}{t+u}{\bf k}^2\right)
d_{\{2,0,0\}}(t,u) 
\right.\nonumber\\ && \left. + 
\left( \frac{3}{2}+\frac{t^2}{t+u}{\bf k}^2\right)
d_{\{0,2,0\}}(t,u) \right]\cdot\frac{1}{t+u}\ 
\label{eq:5.7a}\\
\Omega_4^{(pv)}({\bf k}^2;t,u) &=& +2
(\mbox{\boldmath $\tau$}_1\cdot\mbox{\boldmath $\tau$}_2)
\frac{g_{(\pi\pi)_1}}{m_\pi}
\left(\frac{f_{NN\pi}}{m_\pi}\right)^2\frac{1}{M}\cdot 
\left[ \frac{u}{t+u} d_{\{2,0,0\}} + \frac{t}{t+u} d_{\{0,2,0\}}\right]\ .
\label{eq:5.7b} \end{eqnarray}
\end{subequations}

\noindent (iii)\ $(\pi\sigma)$:
\begin{subequations}
\begin{eqnarray}
\Omega_2^{(pv)}({\bf k}^2;t,u) &=& -
(\mbox{\boldmath $\tau$}_1\cdot\mbox{\boldmath $\tau$}_2)
\frac{g_{(\pi\sigma)_1}}{m_\pi^2}
\frac{f_{NN\pi}}{m_\pi}\frac{g_{NN\sigma}}{M}\cdot d_{\{1,1,1\}}(t,u)
\cdot\left\{ 1+\frac{1}{3}\frac{t^2-tu}{t+u}{\bf k}^2\right\}\frac{1}{t+u}\ .
\label{eq:5.8a}\\
\Omega_3^{(pv)}({\bf k}^2;t,u) &=& -
(\mbox{\boldmath $\tau$}_1\cdot\mbox{\boldmath $\tau$}_2)
\frac{g_{(\pi\sigma)_1}}{m_\pi^2}
\frac{f_{NN\pi}}{m_\pi}\frac{g_{NN\sigma}}{M}\cdot d_{\{1,1,1\}}(t,u)
\cdot\left(\frac{t^2-tu}{t+u}\right)\cdot\frac{1}{t+u}\ .
\label{eq:5.8b} \end{eqnarray}
\end{subequations}

\noindent (iv)\ $(\pi\pi)_0$('derivative'):
\begin{subequations}
\begin{eqnarray}
\Omega_1^{(pv)}({\bf k}^2;t,u) &=& -6 
\left(\frac{g'_{(\pi\pi)_0}}{m_\pi^3}\right)
\left(\frac{f_{NN\pi}}{m_\pi}\right)^2\frac{1}{2M}\cdot d_{1,1,1}(t,u)\cdot 
\left\{\vphantom{\frac{A}{A}} \left(m_1^2-m_2^2\right)^2 \right.
 \nonumber\\ & &        
-\left[3(m_1^2+m_2^2)+\frac{m_1^2(3t^2-u^2)+m_2^2(3u^2-t^2)}{t+u}{\bf k}^2
 \right]\frac{1}{t+u} \nonumber\\ && \left. -
 \left[\left(\frac{t^2+2tu+u^2}{t+u}\right){\bf k}^2 +          
      2tu\left(\frac{t^2+2tu+u^2}{(t+u)^2}\right){\bf k}^4\right]
 \frac{1}{(t+u)^2} \right\}\ ,
\label{eq:5.9a}\\
\Omega_4^{(pv)}({\bf k}^2;t,u) &=& -24
\left(\frac{g'_{(\pi\pi)_0}}{m_\pi^3}\right)
\left(\frac{f_{NN\pi}}{m_\pi}\right)^2\frac{1}{2M}\cdot d_{1,1,1}(t,u)\cdot 
 \nonumber\\ & & \times
 \left\{ \left(m_1^2+m_2^2\right) + 
 \left[\frac{3}{2}+\left(\frac{t^2+tu+u^2}{t+u}\right){\bf k}^2\right]
 \frac{1}{t+u}\right\}\ .
\label{eq:5.9b} \end{eqnarray}
\end{subequations}

The coefficients $\Upsilon_{j,k}^{na,pv}$ defined in (\ref{eq:4.1b}) are 
 tabulated in Tables~\ref{table1}-\ref{table5}.\\

\noindent For $(\pi P)$-exchange, the $1/M_N$ non-adiabatic and pseudo-vector vertex
corrections can be read off from those for $(\pi\sigma)$ and making the same 
adjustments as given already for the adiabatic contributions. In Table~\ref{table6}
the $\Omega^{(pv,ad)}_i$ for $(\pi P)$-pair exchange are given explicitly.

\section{ Partial Wave Analysis}                                
\label{sec:6}
Like the TME-potentials in I, the general form of the MPE-potentials
in momentum space is
\begin{eqnarray}
    \widetilde{V}^{(n)}_{j}({\bf k}) &=&     
 \int_{t_0}^\infty\!\!dt \int_{u_0}^\infty\!\! du\ 
 \left\{w^{(n)}_{\{p_1,p_2,p_3\}}(t,u)\ 
\exp\left[-\left(\frac{tu}{t+u}\right){\bf k}^2\right]\   
 \Omega^{(n)}_j({\bf k}^2;t,u)\right\}\ ,
\label{eq:6.1} \end{eqnarray}
where  
\[
  w^{(n)}_{\{p_1,p_2,p_3\}}(t,u) = w_0(t,u)\ d^{(n)}_{\{p_1,p_2,p_3\}}(t,u)\ .
\]
Therefore, the partial wave analysis runs along the same lines as described in 
sections VI and VII of paper I for the TME-potentials. We refer the reader to
this paper for the details.

\end{widetext}
\section{ ESC-model, Results}                                  
\label{sec:7} 
The momentum space formulas for the potentials of this paper and paper I 
have been checked numerically. This is done by solving the Lippmann-Schwinger
equation and comparing the phase shifts with those obtained by solving
the Schr\"{o}dinger equation using the x-space equivalent of the potentials.

After the completion of the p-space formalism we performed a $\chi^2$-fit
with the ESC-model 
to the 1993 Nijmegen representation of the $\chi^2$-hypersurface of the 
$NN$ scattering data below $T_{lab}=350$ MeV \cite{Sto93}.

\begin{figure}
\resizebox{8.cm}{11.43cm}
 {\includegraphics[50,50][554,770]{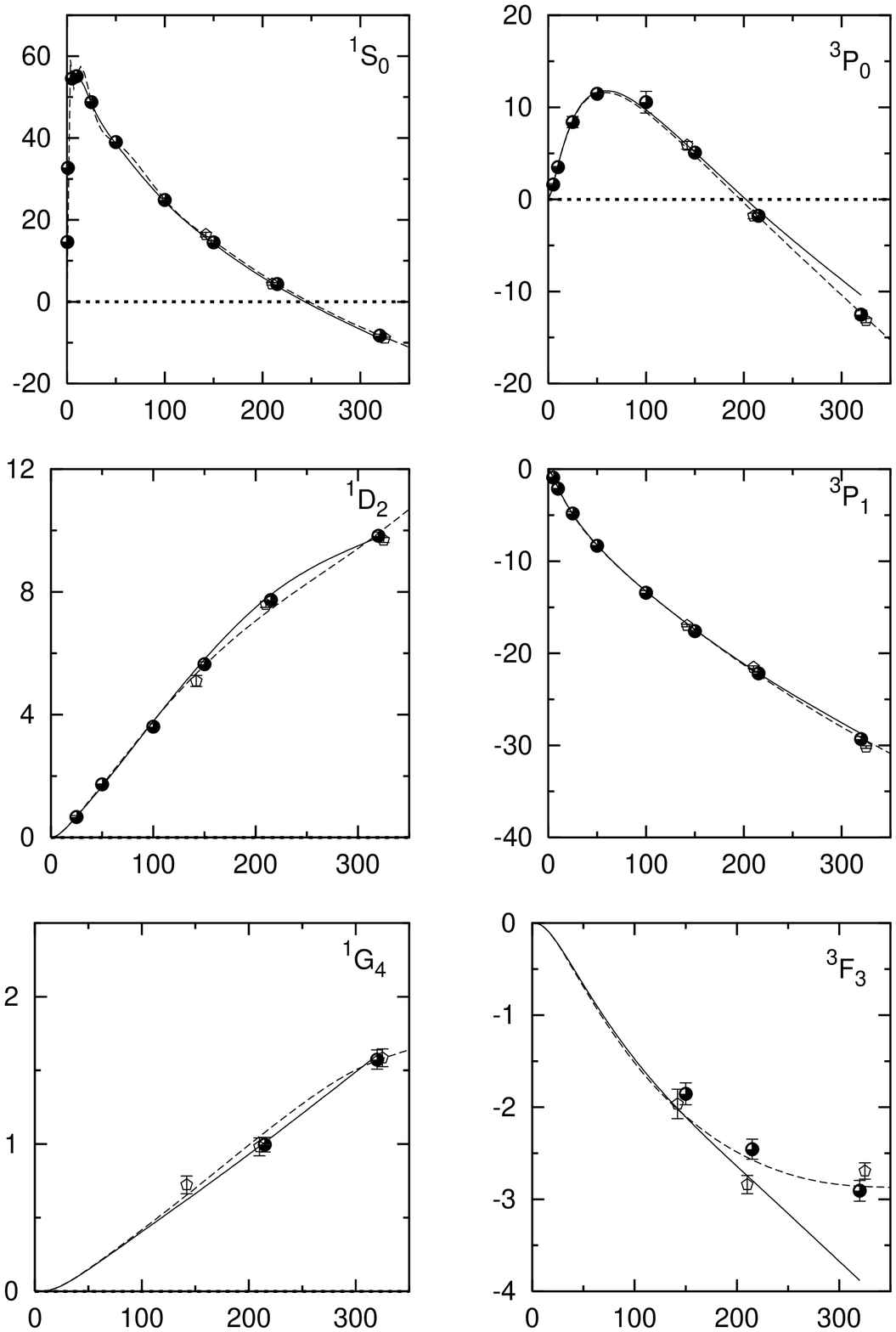}}
\caption{Solid line: proton-proton $I=1$ phase shifts for the ESC-model. 
 The dashed line: the m.e. phases of the Nijmegen93 PW-analysis \cite{Sto93}. 
 The black dots:  the s.e. phases of the Nijmegen93 PW-analysis.
 The diamonds:  Bugg s.e. \cite{Bugg92}.}
\label{ppi1.fig}
 \end{figure}

\begin{figure}   
\resizebox{8.cm}{11.43cm}
 {\includegraphics[50,50][554,770]{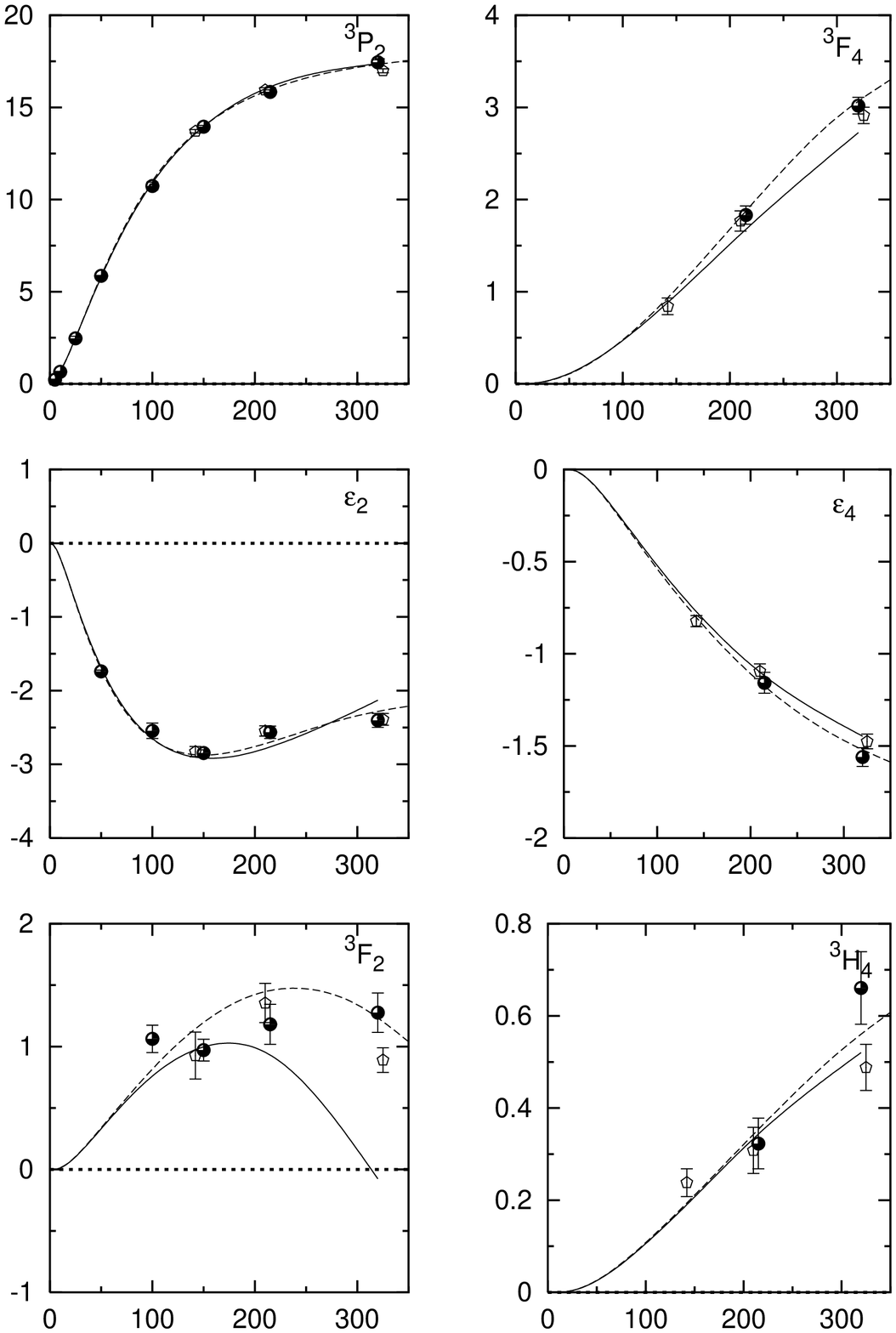}}
\caption{Solid line: proton-proton $I=1$ phase shifts for the ESC-model. 
 The dashed line: the m.e. phases of the Nijmegen93 PW-analysis \cite{Sto93}. 
 The black dots:  the s.e. phases of the Nijmegen93 PW-analysis.
 The diamonds:  Bugg s.e. \cite{Bugg92}.}
\label{ppi1c.fig}
 \end{figure}

\begin{figure}   
\resizebox{8.cm}{7.7cm}
 {\includegraphics[50,280][554,770]{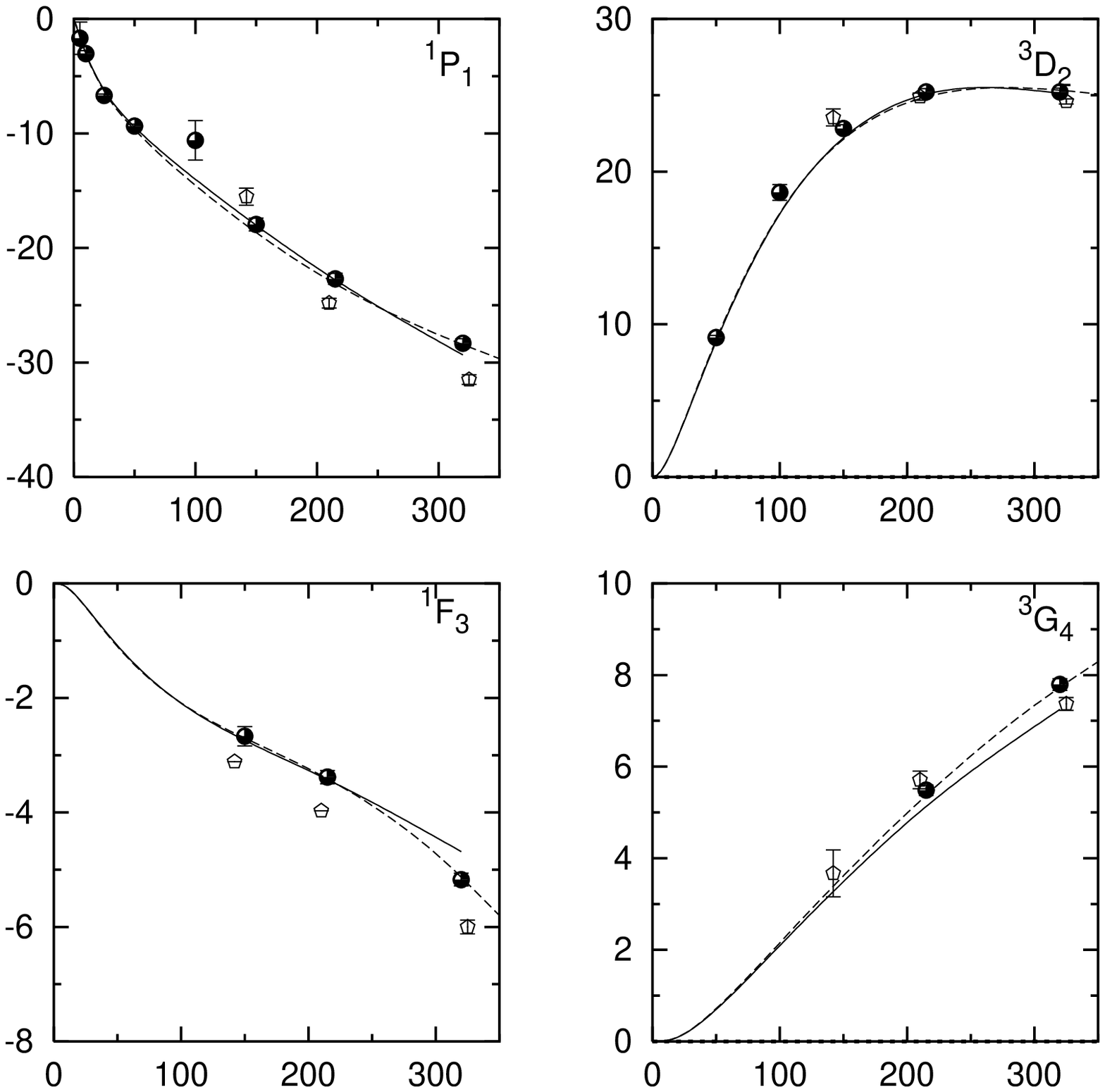}}
\caption{Solid line: neutron-proton $I=0$ phase shifts for the ESC-model. 
 The dashed line: the m.e. phases of the Nijmegen93 PW-analysis \cite{Sto93}. 
 The black dots:  the s.e. phases of the Nijmegen93 PW-analysis.
 The diamonds:  Bugg s.e. \cite{Bugg92}.}
\label{npi0.fig}
 \end{figure}

\begin{figure}   
\resizebox{8.cm}{11.43cm}
 {\includegraphics[50,50][554,770]{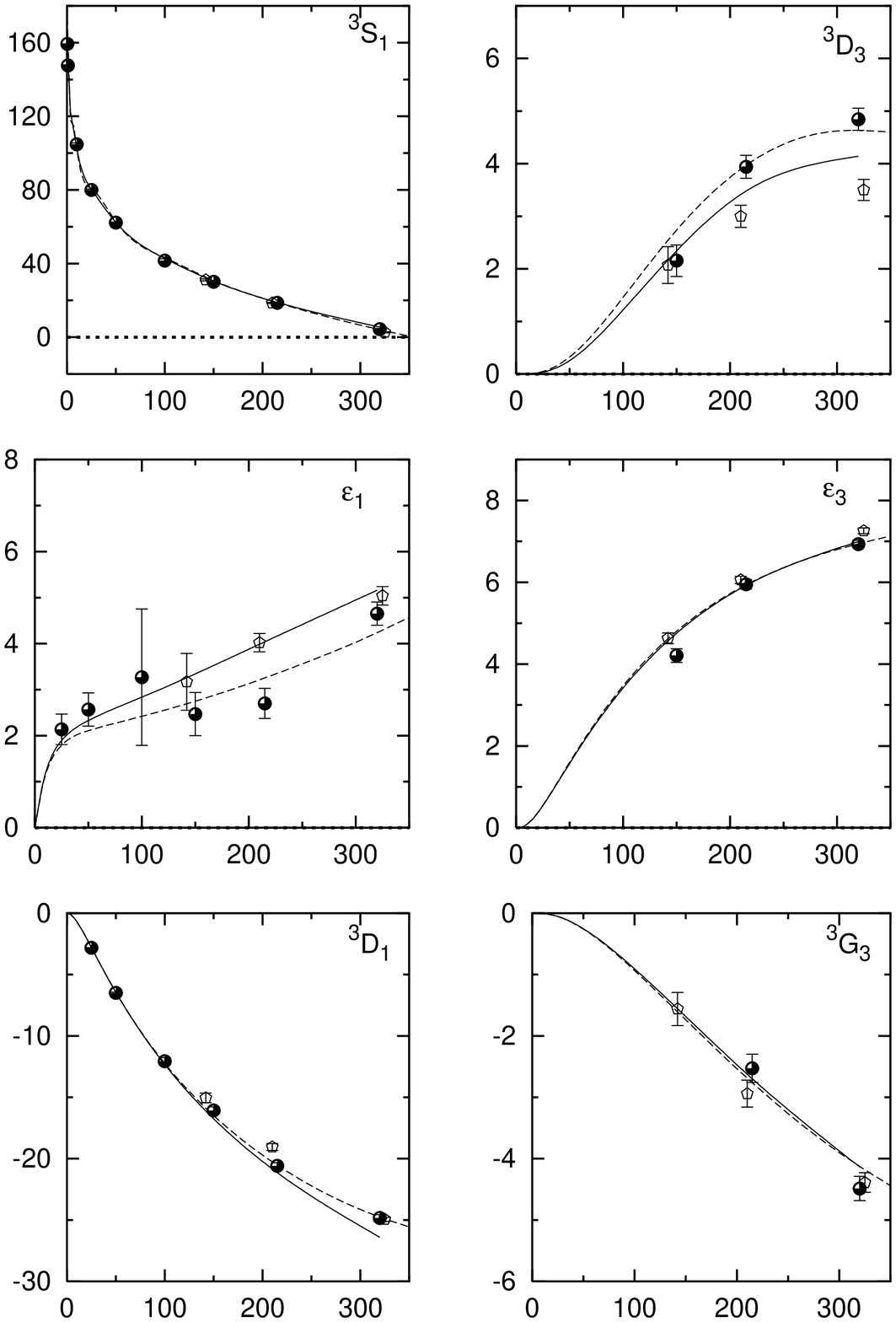}}
\caption{Solid line: neutron-proton $I=0$ phase shifts for the ESC-model. 
 The dashed line: the m.e. phases of the Nijmegen93 PW-analysis \cite{Sto93}. 
 The black dots:  the s.e. phases of the Nijmegen93 PW-analysis.
 The diamonds:  Bugg s.e. \cite{Bugg92}.}
\label{npi0c.fig}
 \end{figure}

This fitting was executed in x-space using the equivalent x-space potentials.
The reason for this is the much faster evaluation of the ESC-model in 
x-space. We obtained a $\chi^2/Ndata =1.15$. The phase shifts are shown in
Fig.s~\ref{ppi1.fig}-\ref{npi0c.fig}.  
In Table~\ref{tab.chidistr} the results are
shown for the ten energy bins, where we compare the results from the updated
partial-wave analysis with the ESC potentials.

In Table~\ref{tab.gobe} we show the OBE-coupling constants and the 
gaussian cut-off's $\Lambda$. The used  $\alpha =: F/(F+D)$-ratio's 
for the OBE-couplings are:
pseudo-scalar mesons $\alpha_{pv}=0.355$, 
vectormesons $\alpha_V^e=1.0, \alpha_V^m=0.275$, 
and scalar-mesons $\alpha_S=0.914$, which is computed using the physical 
$S^* =: f_0(993)$ coupling etc..
In Table~\ref{tab.gpair} we show the MPE-coupling constants.        
The used  $\alpha =: F/(F+D)$-ratio's for the MPE-couplings are:
$(\pi\eta)$ etc. and $(\pi\omega)$ pairs $\alpha(\{8_s\})=1.0$, 
$(\pi\pi)_1$ etc. pairs $\alpha_V^e(\{8\}_a)=0.4, \alpha_V^m(\{8\}_a)=0.335$, 
$(\pi\rho)_1$ etc. pairs $\alpha_A(\{8\}_a)=0.335$. 

\begin{table}[h]
\caption{Meson parameters of the fitted ESC-model. Phases are shown in
         Figs.~\protect\ref{ppi1.fig} to \protect\ref{npi0c.fig}.
         Coupling constants are at ${\bf k}^{2}=0$. 
         An asterisk denotes that the coupling constant is not searched,
         but constrained via $SU(3)$ are simply put to some value used in     
         previous work.}
\begin{ruledtabular}
\begin{tabular}{ccrrc}
meson & mass (MeV) & $g/\sqrt{4\pi}$ & $f/\sqrt{4\pi}$ & $\Lambda$ (MeV) \\
\colrule
 $\pi$         &  138.04 &           & 0.2663   &    950.69  \\
 $\eta$        &  547.45 &           & 0.1461$^{\ast}$ &     ,,    \\
 $\eta'$       &  957.75 &           & 0.1789$^{\ast}$ &      ,,    \\
 $\rho$        &  768.10 &  0.2700   & 3.6378   &    688.20  \\
 $\phi$        & 1019.41 &--1.4717$^{\ast}$  & 0.0149$^{\ast}$ & ,, \\
 $\omega$      &  781.95 &  2.6862   & 0.3255   &      ,,    \\
 $a_{0}$       &  982.70 &  0.9851   &          &    734.25  \\
 $f_{0}$       &  974.10 &--0.7998   &          &      ,,    \\
 $\varepsilon$ &  760.00 &  3.7554   &          &      ,,    \\
 $A_{2}$       &  309.10 &--0.4317   &          &            \\
 Pomeron       &  309.10 &  2.5514   &          &
\end{tabular}
\end{ruledtabular}
\label{tab.gobe}   
\end{table}
 
\begin{table}[h]
\caption{Pair-meson coupling constants employed in the MPE-potentials.     
         Coupling constants are at ${\bf k}^{2}=0$.}
\begin{ruledtabular}
\begin{tabular}{cclcc}
 $J^{PC}$ & $SU(3)$-irrep & $(\alpha\beta)$  & $g/4\pi$  & $f/4\pi$ \\
\colrule
 $0^{++}$ & $\{1\}$  & $(\pi\pi)_{0}$   &  0.1567 &         \\
 $0^{++}$ & ,,       & $(\sigma\sigma)$ &  ---    &         \\
 $0^{++}$ &$\{8\}_s$ & $(\pi\eta)$      &--0.2946 &         \\
 $0^{++}$ &          & $(\pi\eta')$     &  ---    &         \\
 $1^{--}$ &$\{8\}_a$ & $(\pi\pi)_{1}$   &  0.1093 &--0.2050 \\
 $1^{++}$ & ,,       & $(\pi\rho)_{1}$  &  0.6950 &          \\
 $1^{++}$ & ,,       & $(\pi\sigma)$    &  0.0140 &          \\
 $1^{++}$ & ,,       & $(\pi P)$        &--0.1604 &          \\
 $1^{+-}$ &$\{8\}_s$ & $(\pi\omega)$    &--0.1081 &          \\
\end{tabular}
\end{ruledtabular}
\label{tab.gpair}
\end{table}
 
\begin{table}[h]
\caption{$\chi^2$ and $\chi^2$ per datum at the ten energy bins for the    
 Nijmegen93 Partial-Wave-Analysis. $N_{data}$ lists the number of data
 within each energy bin. The bottom line gives the results for the 
 total $0-350$ MeV interval.
 The $\chi^{2}$-access for the ESC model is denoted    
 by  $\Delta\chi^{2}$ and $\Delta\hat{\chi}^{2}$, respectively.}  
\begin{ruledtabular}
\begin{tabular}{crrrrrr} & & & & & \\
 $T_{\rm lab}$ & $\sharp$ data & $\chi_{0}^{2}$\hspace*{5.5mm}&
 $\Delta\chi^{2}$&$\hat{\chi}_{0}^{2}$\hspace*{3mm}&
 $\Delta\hat{\chi}^{2}$ \\ &&&&& \\ \hline
0.383 & 144 & 137.5549 & 21.3 & 0.960 & 0.148  \\
  1   &  68 &  38.0187 & 55.7 & 0.560 & 0.819  \\
  5   & 103 &  82.2257 & 13.0 & 0.800 & 0.127  \\
  10  & 209 & 257.9946 & 78.1 & 1.234 & 0.269  \\
  25  & 352 & 272.1971 & 44.3 & 0.773 & 0.126  \\
  50  & 572 & 547.6727 &137.4 & 0.957 & 0.240  \\
  100 & 399 & 382.4493 & 27.6 & 0.959 & 0.069  \\
  150 & 676 & 673.0548 & 82.9 & 0.996 & 0.123  \\
  215 & 756 & 754.5248 &108.0 & 0.998 & 0.143  \\
  320 & 954 & 945.3772 &305.0 & 0.991 & 0.320  \\ \hline
      &    &     &     &     &    \\
Total &4233&4091.122& 864.2 &0.948 &0.201  \\
      &    &     &     &     &     \\
\end{tabular}
\end{ruledtabular}
\label{tab.chidistr} 
\end{table}

We emphasize that we use the single-energy (s.e.) phases and $\chi^2$-surface 
\cite{Klo93}
only as a means to fit the NN-data. As stressed in \cite{Sto93} the 
Nijmegen s.e. phases have not much significance. The significant phases 
are the multi-energy (m.e.) ones, see the dashed lines in the figures.
One notices that the central value of the s.e. phases do not correspond
to the m.e. phases in general,
illustrating that there has been a certain amount
of noice fitting in the s.e. PW-analysis, see e.g. $\epsilon_1$ and $^1P_1$ 
at $T_{lab}=100$ MeV.
The m.e. PW-analysis reaches $\chi^2/N_{data}=0.99$, using 
 39 phenomenological parameters plus normalization parameters.
The related phenomenological PW-potentials NijmI,II and Reid93 \cite{SKTS94},
with respectively 41, 47, and 50 parameters, all with $\chi^2/Ndata=1.03$.
This should be compared to the ESC-model, which has $\chi^2/N_{data}=1.15$
using 20 parameters. These are 9 meson-nucleon-nucleon couplings,
8 meson-pair-nucleon-nucleon couplings, and 3 gaussian cut-off parameters.
{}From the figures it is obvious that the ESC-model deviates from the m.e.
PW-analysis at the highest energy in particular. If we evalute the $\chi^2$
for the first 9 energies only, we obtain $\chi^2/N_{data} = 1.10$.

We mentioned that we do not include negative energy state contributions.
It is assumed that a strong pair suppression is operative at low energies
in view of the composite nature of the nucleons. This leaves us for the
pseudo-scalar mesons with two essential equivalent interactions: the 
direct and the derivative one. In expanding the $NN\pi$- etc. vertex in   
$1/M_N$ these two interactions differ in the $1/M^2_N$-terms, see \cite{RS96a} 
equations (3.4) and (3.5).
Here, we prefer to cancel these $1/M_N^2$ terms by taking 
\begin{equation}
 {\cal H}_{ps} = \frac{1}{2}\left[g_{NN\pi} \bar{\psi}i\gamma_5
 \mbox{\boldmath $\tau$}\psi\cdot\mbox{\boldmath $\pi$} + (f_{NN\pi}/m_\pi)
 \gamma_\mu\gamma_5\mbox{\boldmath $\tau$}\psi\cdot
 \partial^\mu\mbox{\boldmath $\pi$}\right]\ ,
\end{equation}
where $g_{NN\pi}= (2M_N/m_\pi) f_{NN\pi}$.

As for the OBE-couplings, one notices that $G_E = g_{\rho NN}$ is small, but
$G_M = g_\rho + f_\rho$ is okay.
One possible explanation would be that part of the $\rho$-exchange is 
replaced by the 2-pair $(\pi\pi)_1$-exchange, 
which has identical quantum numbers. This still leaves room for 
the interpretation of the 1-pair $(\pi\pi)_1$-exchange as a form factor 
correction.   
Another interesting possibility is that leaving out the tensor 
mesons $a_2(1320),f_2(1270),f_2(1520)$ affects the vector meson couplings.
This can be seen as follows. At high energies and low to moderate 
momentum transfer there is a strong cancellation between the vector and
tensor exchange: $(\rho-a_2)$- and $(\omega-f_2)$-cancellation \cite{Arn65}.
This is called Exchange-Degeneracy (EXD).
Indeed, by changing $g_\rho/\sqrt{4\pi}=0.3$ to $g_\rho=0.75/\sqrt{4\pi}$
one can cancel the change in the $\rho$-exchange potential by the inclusion
of $a_2$-exchange rather completely. 
The inclusion of mesons with a mass $\ge 1$ GeV$/c^2$, like the axial and
tensor mesons we leave as a future project.

Unlike in \cite{RS96a,RS96b}, we did not fix pair couplings using
a theoretical model, based on heavy-meson saturation and chiral-symmetry.
So, in addition to the 14 parameters used in \cite{RS96a,RS96b} we now have
6 pair-coupling fit parameters. 
In Table~\ref{tab.gpair} the fitted pair-couplings are given.
Note that the $(\pi\pi)_0$-pair coupling gets contributions from the $\{1\}$ and
the $\{8_s\}$ pairs as well, giving in total $g_{(\pi\pi)}=0.10$, which has the
same sign as in \cite{RS96b}. The $f_{(\pi\pi)_1}$-pair coupling has opposite
sign as compared to \cite{RS96b}. In a model with a more complex and realistic
meson-dynamics \cite{SR97} this coupling is predicted as found in the present 
ESC-fit. The $(\pi\rho)_1$-coupling agrees nicely with $A_1$-saturation, see 
\cite{RS96b}. We conclude that the pair-couplings are in general not well
understood, and deserve more study.

The ESC-model described here is fully consistent with $SU(3)$-symmetry. In
Appendix~\ref{app:F.a} we display the full $SU(3)$ contents of the 
pair interaction Hamiltonians. For example $g_{(\pi\rho)_1} = g_{A_8VP}$, and
besides $(\pi\rho)$-pairs one sees also that $(K K^*(I=1)$- and 
$K K^*(I=0)$-pairs contribute to the $NN$ potentials.
All $F/(F+D)$-ratio's are taken fixed with heavy-meson saturation in mind.
The approximation we have made in this paper is to neglect the baryon mass
differences, i.e. we put $m_\Lambda = m_\Sigma = m_N$. This because we
have not yet worked out the formulas for the inclusion of these mass 
differences, which is straightforward in principle.

\section{ Conclusions and Outlook}                             
\label{sec:8} 
The presented ESC-model is very succesfull and flexible in describing the NN-data. 
It can be developed and extended in various ways.
First, we plan to extend the OBE-potentials in momentum space by
including the full OBE-propagator, i.e. 
\begin{equation}
 \frac{1}{\omega^2} \rightarrow \frac{1}{\omega (\omega+a)}\ ,\ 
 a = \frac{1}{M}\left[p_f^2+p_i^2-2p_0^2\right]\ .
\label{eq:8.1} \end{equation}
This includes retardation at the level of the OBE-potentials.
Secondly, one may extend the TME-potentials including besides ps-ps also
the ps-vector, ps-scalar, etc. potentials. Thirdly, the inclusion of the 
axial- and tensor-mesons, which we discussed in connection with EXD.

The momentum space formulation of the ESC-model also suggests       
a covariant formulation. Consider an Effective Field Theory and suppose 
that it allows the Wick-rotation. Then, assuming in Euclidean space  
a Gaussian cut-off, one can use a representation completely
akin to \ref{eq:2.3} etc. For example, this opens the way to analyse the expansion in
loops in the presence of a strong cut-off. Also, one could evaluate the ESC-model
using the Bethe-Salpetyer equation.

The presented ESC-model can be applied  in various ways:
(i) The study of Few-body systems in momentum space,                 
(ii) The study of Meson-Exchange-Current (MEC) corrections,           
(iii) The derivation of 3-body forces consistent with the 2-body forces.
(iv) G-matrix etc. description of Nuclear Matter.


\appendix

\begin{widetext}

\section{ Pair Interaction Hamiltonians}      
\label{app:AA}
The pair hamiltonians are
\begin{subequations}
\begin{eqnarray}
 J^{PC} = 0^{++} & : & {\cal H}_S = (\bar{\psi}'\psi')\left\{
 g_{\pi\pi)_0}(\mbox{\boldmath $\pi$}\cdot\mbox{\boldmath $\pi$}) +   
 g_{\sigma\sigma} \sigma^2\right\}/m_\pi + \nonumber\\
 && \nonumber\\
 && \hspace{1cm} g'_{(\pi\pi)_0}(\bar{\psi}'\psi)
 (\partial_\mu\mbox{\boldmath $\pi$}\cdot
 \partial^\mu\mbox{\boldmath $\pi$})/     
 m_\pi^3\ , \label{appAA.1a}\\
 && \nonumber\\
 J^{PC} = 1^{--} & : & {\cal H}_V = \left[g_{(\pi\pi)_1}
 \bar{\psi}'\gamma_\mu\mbox{\boldmath $\tau$}\psi'
 -\frac{f_{(\pi\pi)_1}}{2M}
 \bar{\psi}'\sigma_{\mu\nu}\mbox{\boldmath $\tau$}\psi'\partial^\nu\right]
 (\mbox{\boldmath $\pi$}\times\partial^\mu\mbox{\boldmath $\pi$})/m_\pi^2\ ,    
 \label{appAA.1b}\\
 && \nonumber\\
 J^{PC} = 1^{++} & : & {\cal H}_A = g_{(\pi\rho)_1}
 \bar{\psi}'\gamma_\mu\gamma_5\mbox{\boldmath $\tau$}\psi'
 (\mbox{\boldmath $\pi$}\times\mbox{\boldmath $\rho$}^\mu)/m_\pi + \nonumber\\
 && \nonumber\\
                 &   & \hspace{1cm}  g_{(\pi\sigma)}
 \bar{\psi}'\gamma_\mu\gamma_5\mbox{\boldmath $\tau$}\psi'
 \left(\sigma\partial^\mu\mbox{\boldmath $\pi$}
 -\mbox{\boldmath $\pi$}\partial^\mu\sigma\right)/m_\pi^2\ ,
 \label{appAA.1c}\\
 && \nonumber\\
 J^{PC} = 1^{+-} & : & {\cal H}_B = g_{(\pi\rho)_0}
 \bar{\psi}'\sigma_{\mu\nu}\gamma_5\psi' \partial^\nu
 (\mbox{\boldmath $\pi$}\cdot\mbox{\boldmath $\rho$})/m_\pi^2 + \nonumber\\
 && \nonumber\\
 && \hspace{1cm} g_{\pi\omega}
 \bar{\psi}'\sigma_{\mu\nu}\gamma_5\mbox{\boldmath $\tau$}\psi'
 \partial^\nu(\mbox{\boldmath $\pi$}\cdot\omega^\mu)/m_\pi^2\ .    
 \label{appAA.1d}
\end{eqnarray}
\end{subequations}

\section{ $\lambda$-representations}      
\label{app:A}
The following $\lambda$-representations \cite{Rij91} are exploited:
\begin{subequations}
\begin{eqnarray}
 D_{1,0,0}(\omega_1,\omega_2) &=& \frac{1}{\omega_1} = 
\frac{2}{\pi} \int_0^\infty 
\frac{d\lambda}{\omega_1^2+\lambda^2}\ , \label{appA.1a}\\
 D_{0,1,0}(\omega_1,\omega_2) &=& \frac{1}{\omega_1} = 
\frac{2}{\pi} \int_0^\infty 
\frac{d\lambda}{\omega_2^2+\lambda^2}\ , \label{appA.1b}\\
 D_{0,0,1}(\omega_1,\omega_2) &=& \frac{1}{\omega_1+\omega_2} =
\frac{2}{\pi} \int_0^\infty 
\frac{\lambda^2 d\lambda}{(\omega_1^2+\lambda^2)(\omega_2^2+\lambda^2)}\ 
 , \label{appA.1c}\\
 D_{1,1,1}(\omega_1,\omega_2) &=& 
\frac{1}{\omega_1\omega_2(\omega_1+\omega_2)} =
\frac{2}{\pi} \int_0^\infty 
\frac{d\lambda}{(\omega_1^2+\lambda^2)(\omega_2^2+\lambda^2)}\ 
 , \label{appA.1d}
\end{eqnarray}
\end{subequations}

A special combination occurs in non-adiabatic terms. Here, see Table~\ref{D1pair},
occurs 
\begin{eqnarray}
 D_{na}^{(1)}(\omega_1,\omega_2) &=& \frac{1}{\omega_1^2\omega_2^2}
 \left[\frac{1}{\omega_1}+\frac{1}{\omega_2}-\frac{1}{\omega_1+\omega_2}\right] =
 \nonumber\\
 &=& \frac{2}{\pi}\int_0^\infty \frac{d\lambda}{\lambda^2}
 \left[ \frac{1}{\omega_1^2\omega_2^2}-
 \frac{1}{(\omega_1^2+\lambda^2)(\omega_2^2+\lambda^2)}\right]\ .
\label{appA.2} \end{eqnarray}
Notice that the denominator $D_{na}^{(1)} = 2 D_{//}$, see \cite{RS96a}.
The corresponding $d_{na}(t,u)$ is, see paper I, section IV.A,
\begin{equation}
 d_{na}(t,u) = \frac{2}{\pi}\int_0^\infty\frac{d\lambda}{\lambda^2}\left[
 1-e^{-(t+u)\lambda^2}\right]=\frac{2}{\sqrt{\pi}}\sqrt{t+u}\ .
\label{appA.3} \end{equation}

\section{ Integration Dictionary}         
\label{app:D}
In this appendix we give a dictionary for the evaluation of the momentum
integrals that occur in the matrix elements of the TME-potentials.
The results of the $d^3\Delta$-integration are given apart from a factor
$(4\pi a)^{-3/2}\ (a= t+u)$, common to all integrals. Using the results given in
Appendix B of paper I, one obtains:

 \noindent (i)\ For the operators $\widetilde{O}^{(1)}_{\alpha\beta,p}$,
 and the operators $\widetilde{O}^{(2)}_{\alpha\beta,p}$:
\begin{subequations}
\begin{eqnarray}
 && a.\ \left({\bf k}_1\cdot{\bf k}_2\right)
 = \mbox{\boldmath $\Delta$}\cdot{\bf k}
 -\mbox{\boldmath $\Delta$}^2 \Rightarrow
\frac{1}{2}\left\{-3+2\left(\frac{tu}{t+u}\right){\bf k}^2\right\}\cdot
\frac{1}{t+u}\ ,
\label{AppD.1a} \\ && \nonumber\\ 
 && b.\ \left[\mbox{\boldmath $\sigma$}_1\cdot{\bf k}_1\times{\bf k}_2\right] 
 \left[\mbox{\boldmath $\sigma$}_2\cdot{\bf k}_1\times{\bf k}_2\right] =
 \left[\mbox{\boldmath $\sigma$}_1\cdot\mbox{\boldmath $\Delta$}\times{\bf k}\right]
 \left[\mbox{\boldmath $\sigma$}_2\cdot\mbox{\boldmath $\Delta$}\times{\bf k}\right]
 \nonumber\\ && \hspace{5mm}\Rightarrow \frac{1}{2}\left\{\frac{2}{3}
 \left(\mbox{\boldmath $\sigma$}_1\cdot\mbox{\boldmath $\sigma$}_2\right)\
{\bf k}^2\ - \left[
 \left(\mbox{\boldmath $\sigma$}_1\cdot{\bf k}\right)\
 \left(\mbox{\boldmath $\sigma$}_2\cdot{\bf k}\right) - \frac{1}{3}
 \left(\mbox{\boldmath $\sigma$}_1\cdot\mbox{\boldmath $\sigma$}_2\right)\
 {\bf k}^2\right]\right\}\cdot \frac{1}{t+u}\ ,              
\label{AppD.1b} \\  && \nonumber\\   
 && c.\ [\left(\mbox{\boldmath $\sigma$}_1+\mbox{\boldmath $\sigma$}_2\right)\cdot
 {\bf k}_1\times{\bf k}_2]\ {\bf q}\cdot\left({\bf k}_1-{\bf k}_2\right) =
 [\left(\mbox{\boldmath $\sigma$}_1+\mbox{\boldmath $\sigma$}_2\right)\cdot
 \mbox{\boldmath $\Delta$}\times{\bf k}]\ 
 {\bf q}\cdot\left(\mbox{\boldmath $2\Delta$}-{\bf k}\right) 
\nonumber\\ && \hspace{5mm}\Rightarrow 
 [\left(\mbox{\boldmath $\sigma$}_1+\mbox{\boldmath $\sigma$}_2\right)\cdot
 {\bf q}\times{\bf k}]\cdot\frac{1}{t+u}\ ,
\label{AppD.1c} \\  && \nonumber\\ 
 && d.\ (\mbox{\boldmath $\sigma$}_1\cdot{\bf k}_1\                            
 \mbox{\boldmath $\sigma$}_2\cdot{\bf k}_2) + (\mbox{\boldmath $\sigma$}_1\cdot{\bf k}_2\           
 \mbox{\boldmath $\sigma$}_2\cdot{\bf k}_1)                                
\nonumber\\ && \hspace{5mm}\Rightarrow 
 -\left\{ \mbox{\boldmath $\sigma$}_1\cdot\mbox{\boldmath $\sigma$}_2 -
 \frac{2tu}{t+u} \left(\mbox{\boldmath $\sigma$}_1\cdot{\bf k}\right)\
 \left(\mbox{\boldmath $\sigma$}_2\cdot{\bf k}\right) \right\}\cdot\frac{1}{t+u}\ .
\label{AppD.1d}\\                  
 && e.\ (\mbox{\boldmath $\sigma$}_1\cdot{\bf k}_1)                             
 (\mbox{\boldmath $\sigma$}_2\cdot{\bf k}_1) =                              
 (\mbox{\boldmath $\sigma$}_1\cdot\mbox{\boldmath $\Delta$})             
 (\mbox{\boldmath $\sigma$}_2\cdot\mbox{\boldmath $\Delta$})             
\nonumber\\ && \hspace{5mm}\Rightarrow 
 \frac{1}{2}\left\{
 \mbox{\boldmath $\sigma$}_1\cdot\mbox{\boldmath $\sigma$}_2 +
 \frac{2u^2}{t+u} \left(\mbox{\boldmath $\sigma$}_1\cdot{\bf k}\right)\
 \left(\mbox{\boldmath $\sigma$}_2\cdot{\bf k}\right) \right\}\cdot\frac{1}{t+u}\ ,
\label{AppD.1e}\\                 
 && f.\ \mbox{\boldmath $\sigma$}_1\cdot({\bf k}_1-{\bf k}_2)\                            
 \mbox{\boldmath $\sigma$}_2\cdot({\bf k}_1-{\bf k}_2) =                              
 \mbox{\boldmath $\sigma$}_1\cdot(2\mbox{\boldmath $\Delta$}-{\bf k})\            
 \mbox{\boldmath $\sigma$}_2\cdot(2\mbox{\boldmath $\Delta$}-{\bf k})             
\nonumber\\ && \hspace{5mm}\Rightarrow 
 \mbox{\boldmath $\sigma$}_1\cdot\mbox{\boldmath $\sigma$}_2 
 \left\{ \frac{2}{t+u}+\frac{1}{3}\left(\frac{t-u}{t+u}\right)^2{\bf k}^2\right\} +
\left(\mbox{\boldmath $\sigma$}_1\cdot{\bf k}\ \mbox{\boldmath $\sigma$}_2\cdot{\bf k} 
 -\frac{1}{3}{\bf k}^2 \mbox{\boldmath $\sigma$}_1\cdot\mbox{\boldmath $\sigma$}_2 
\right)\cdot\left(\frac{t-u}{t+u}\right)^2\ ,
\label{AppD.1f} \end{eqnarray}
\end{subequations}

 \noindent (ii)\ For the $1/M$-correction operators 
 $\widetilde{O}^{(na)}_{\alpha\beta}$ etc., not included in the list 
 (\ref{AppD.1a})-(\ref{AppD.1e}):

\begin{subequations}
\begin{eqnarray}
 && a.\ ({\bf k}_1\cdot{\bf k}_2)^2 = \left(\mbox{\boldmath $\Delta$}\cdot{\bf k}
 -\mbox{\boldmath $\Delta$}^2\right)^2 \nonumber\\ 
 &&\hspace{5mm}\Rightarrow \frac{1}{4}\left\{ 15 + 
 2\left(\frac{t^2-8ut+u^2}{t+u}\right){\bf k}^2 + 4\left(\frac{t^2u^2}{(t+u)^2}\right)
 {\bf k}^4\right\}\cdot\frac{1}{(t+u)^2}\ ,
\label{AppD.2a} \\ && \nonumber\\ 
 && b.\ (\mbox{\boldmath $\sigma$}_1\cdot{\bf k}_2)                             
 (\mbox{\boldmath $\sigma$}_2\cdot{\bf k}_2) =                              
 \mbox{\boldmath $\sigma$}_1\cdot({\bf k}-\mbox{\boldmath $\Delta$})\             
 \mbox{\boldmath $\sigma$}_2\cdot({\bf k}-\mbox{\boldmath $\Delta$})             
\nonumber\\ && \hspace{5mm}\Rightarrow 
 \frac{1}{2}\left\{
 \mbox{\boldmath $\sigma$}_1\cdot\mbox{\boldmath $\sigma$}_2 +
 \frac{2t^2}{t+u} \left(\mbox{\boldmath $\sigma$}_1\cdot{\bf k}\right)\
 \left(\mbox{\boldmath $\sigma$}_2\cdot{\bf k}\right) \right\}\cdot\frac{1}{t+u}\ ,
\label{AppD.2b} \\  && \nonumber\\ 
 && c.\ {\bf k}_1^2 = \mbox{\boldmath $\Delta$}^2 \Rightarrow 
 \left\{\frac{3}{2} + \frac{u^2}{t+u} {\bf k}^2\right\}\cdot\frac{1}{t+u}\ ,
\label{AppD.2c} \\ && \nonumber\\  
 && d.\ {\bf k}_2^2 = \left({\bf k}-\mbox{\boldmath $\Delta$}\right)^2 \Rightarrow 
 \left\{\frac{3}{2} + \frac{t^2}{t+u} {\bf k}^2\right\}\cdot\frac{1}{t+u}\ ,
\label{AppD.2d} \\ && \nonumber\\  
 && e.\ {\bf k}_1^2\ {\bf k}_2^2 = {\bf k}^2 \mbox{\boldmath $\Delta$}^2
 -2{\bf k}\cdot\mbox{\boldmath $\Delta$}\ \mbox{\boldmath $\Delta$}^2
 +\mbox{\boldmath $\Delta$}^4 \nonumber\\
 && \Rightarrow \left[\frac{15}{4}+\frac{1}{2}\frac{(3t^2-4tu+3u^2)}{t+u}{\bf k}^2
 + \frac{t^2u^2}{(t+u)^2} {\bf k}^4\right\}\cdot\frac{1}{(t+u)^2}\ ,
\label{AppD.2e} \\ && \nonumber\\  
 && f.\ ({\bf k}_1\cdot{\bf k}_2)
 (\mbox{\boldmath $\sigma$}_1\cdot{\bf k}_1)                             
 (\mbox{\boldmath $\sigma$}_2\cdot{\bf k}_1) =                              
 \mbox{\boldmath $\Delta$}\cdot({\bf k}-\mbox{\boldmath $\Delta$})
 [\mbox{\boldmath $\sigma$}_1\cdot\mbox{\boldmath $\Delta$}\             
 \mbox{\boldmath $\sigma$}_2\cdot\mbox{\boldmath $\Delta$}]            
\nonumber\\ && \hspace{5mm}\Rightarrow 
 \left\{ -\frac{5}{4}+\frac{1}{2}\frac{tu}{t+u}{\bf k}^2\right\}\cdot\frac{1}{(t+u)^2}
 \mbox{\boldmath $\sigma$}_1\cdot\mbox{\boldmath $\sigma$}_2 \nonumber\\ \hspace{1cm} && +
\left(\mbox{\boldmath $\sigma$}_1\cdot{\bf k}\ \mbox{\boldmath $\sigma$}_2\cdot{\bf k}\right) 
 \left\{ \frac{1}{2}\frac{(2t-5u)u}{t+u}
 +\frac{tu^3}{(t+u)^2}{\bf k}^2\right\}\cdot\frac{1}{(t+u)^2}\ ,
\label{AppD.2f} \\  && \nonumber\\ 
 && g.\ ({\bf k}_1\cdot{\bf k}_2)
 (\mbox{\boldmath $\sigma$}_1\cdot{\bf k}_1)                             
 (\mbox{\boldmath $\sigma$}_2\cdot{\bf k}_2) =                              
 \mbox{\boldmath $\Delta$}\cdot({\bf k}-\mbox{\boldmath $\Delta$})
 [\mbox{\boldmath $\sigma$}_1\cdot\mbox{\boldmath $\Delta$}\             
 \mbox{\boldmath $\sigma$}_2\cdot({\bf k}-\mbox{\boldmath $\Delta$})]           
\nonumber\\ && \hspace{5mm}\Rightarrow 
 \left\{\frac{5}{4}-\frac{1}{2}\frac{tu}{t+u}{\bf k}^2\right\}\cdot\frac{1}{(t+u)^2}
 \mbox{\boldmath $\sigma$}_1\cdot\mbox{\boldmath $\sigma$}_2 \nonumber\\ && \hspace{1cm} +
\left(\mbox{\boldmath $\sigma$}_1\cdot{\bf k}\ \mbox{\boldmath $\sigma$}_2\cdot{\bf k}\right) 
 \left\{ \frac{1}{2} -\frac{7}{2}\frac{tu}{(t+u)^2}
 +\frac{t^2u^2}{(t+u)^3}{\bf k}^2\right\}\cdot\frac{1}{t+u}\ .
\label{AppD.2g} \end{eqnarray}
\end{subequations}

\section{Derivative Scalar-Pair Potentials}
\label{app:E}
As pointed out by Ko and Rudaz \cite{Ko94} besides the most simple lagrangian
for $\sigma$-decay ${\cal L}^{(0)}_{\sigma\pi\pi} = g_{\sigma\pi\pi} \sigma
\mbox{\boldmath $\pi$}\cdot\mbox{\boldmath $\pi$}$ also the lagrangian 
$\sigma$-decay ${\cal L}^{(1)}_{\sigma\pi\pi} = g'_{\sigma\pi\pi} \sigma
\partial_\mu\mbox{\boldmath $\pi$}\cdot\partial^\mu\mbox{\boldmath $\pi$}$  
appears in the linear $\sigma$-model. The latter is useful in keeping the 
scalar meson width's within reasonable bounds as the scalar mass increases.
Also, derivative couplings to baryons were considered in the context of an
$SU_f(3)$ generalization in \cite{SR97}. In the $(NN2\pi)$ effective
field-theory lagrangian \cite{Ordonez92} the $NN$-interaction lagrangian, i.e.
the NLO-terms, for the pion-pairs reads
\begin{equation}
{\cal L}^{(1)} = 
 -\bar{\psi}\left[8c_1D^{-1}m_\pi^2\frac{\mbox{\boldmath $\pi$}}
{F_\pi^2}-4c_3{\bf D}_\mu\cdot{\bf D}^\mu + 2c_4\sigma_{\mu\nu}
\mbox{\boldmath $\tau$}\cdot{\bf D}^\mu\times{\bf D}^\nu\right] \psi\ ,
\label{AppE.1} \end{equation}
where $D=1+\mbox{\boldmath $\pi$}^2/f_\pi^2$ and ${\bf D}_\mu = D^{-1}
\partial_\mu\mbox{\boldmath $\pi$}/F_\pi$, with $F_\pi=2 f_\pi = 185$ MeV.
The correspondence with the pair terms treated in this paper is that 
$c_1 \sim g_{(\pi\pi)_0}$, and $c_3 \sim g'_{(\pi\pi)_0}$. 
The $c_4$-term has been considered in \cite{SR97}, but not in this paper.
The 'derivative' hamiltonian to lowest order in the 
$\mbox{\boldmath $\pi$}$ reads
\begin{equation}
 {\cal H}_{S'} = g'_{(\pi\pi)_0}(\bar{\psi}'\psi)
 (\partial_\mu\mbox{\boldmath $\pi$}\cdot
 \partial^\mu\mbox{\boldmath $\pi$})/m_\pi^3\ . 
\label{AppE.2} \end{equation}

\subsection{ Adiabatic potentials}           
\label{app:E.a}
For the 1-pair graph's in equation (\ref{eq:3.1}) 
\begin{eqnarray}
 && \tilde{O}^{(1)}_{\alpha\beta,p}({\bf k}_{1},{\bf k}_2) \Rightarrow
 \tilde{O}^{(S')}_{\alpha\beta,p} \tilde{O}^{(2PS)}_{\alpha\beta}\ \ , \ \
 \tilde{O}^{(2PS)}_{\alpha\beta}=-\left(\frac{f_{NN\pi}}{m_\pi}\right)^2
 ({\bf k}_1\cdot{\bf k}_2-i\mbox{\boldmath $\sigma$}\cdot{\bf k}_1\times{\bf k}_2)\,
 \nonumber\\ &&
 \tilde{O}^{(S')}_{\alpha\beta,p} = 2\frac{g'_{(\pi\pi)_0}}{m_\pi^3}\left(     
 \pm \omega_1\omega_2+{\bf k}_1\cdot{\bf k}_2\right)\ .
\label{AppE.3} \end{eqnarray}
Here, for $p=a,c$ the $(-)$-sign and for $p=b$ the $(+)$-sign applies. 
Obviously, $\alpha=\beta=\pi$ in (\ref{AppE.3}). All other
quantities in (\ref{eq:3.1}) are the same as for pion-pair without derivatives.
Here, and in the rest of this appendix, we absorb the $g^{(n)}(\alpha,\beta)$-factor
in (\ref{eq:3.1}) into the definition of the $O$-operators.\\
Evaluation of the p-sum and including the mirror graph's, one gets collecting
all terms and selecting the contributions symmetric in $1 \leftrightarrow 2$ the
matrix element
\begin{eqnarray}
 && \sum_p \tilde{O}^{(1)}_{\alpha\beta,p}({\bf k}_{1},{\bf k}_2) 
 D_p^{(1)}(\omega_1,\omega_2) = - 2
 \frac{g'_{(\pi\pi)_0}}{m_\pi^3}\left(\frac{f_{NN\pi}}{m_\pi}\right)^2\cdot
 \nonumber \\ && \times ({\bf k}_1\cdot{\bf k}_2) \left\{
 {\bf k}_1\cdot{\bf k}_2-i\mbox{\boldmath $\sigma$}\cdot{\bf k}_1\times{\bf k}_2
 \right\}\frac{1}{\omega_1^2\omega_2^2}\ .
\label{AppE.4} \end{eqnarray}
For the 2-pair graph's in equation (\ref{eq:3.1}) one has
\begin{eqnarray}
 && \tilde{O}^{(2)}_{\alpha\beta,p}({\bf k}_{1},{\bf k}_2) 
 D^{(2)}(\omega_1,\omega_2) = -\frac{1}{2\omega_1\omega_2(\omega_1+\omega_2)}
 \left(-\omega_1\omega_2+{\bf k}_1\cdot{\bf k}_2\right)^2\ .
\label{AppE.5} \end{eqnarray}

Using the expressions in this appendix  we obtain  
in p-space the adiabatic 'derivative' $(\pi\pi)_0$-exchange potentials
 the 1-pair exchange and 2-pair graphs give
\begin{subequations}
\begin{eqnarray}
\Omega^{(1),ad}_1({\bf k}^2;t,u) &=& -12
 \left(\frac{g'_{(\pi\pi)_0}}{m_\pi^3}\right)
 \left(\frac{f_{NN\pi}}{m_\pi}\right)^2\cdot 
 d_{2,2,0}(t,u)\cdot \nonumber\\ && \times
\left[\frac{15}{4}+\frac{1}{2}\frac{t^2-8tu+u^2}{t+u}{\bf k}^2
 + \frac{t^2u^2}{(t+u)^2}{\bf k}^4\right] 
\cdot\frac{1}{(t+u)^2}\ , \label{AppE.6a} \\
\Omega^{(2),ad}_1({\bf k}^2;t,u) &=& -6
 \left(\frac{g'_{(\pi\pi)_0}}{m_\pi^3}\right)^2\cdot
\left\{\left[\frac{15}{4}+\frac{t^2-3tu+u^2}{t+u}{\bf k}^2+
\frac{t^2u^2}{(t+u)^2}{\bf k}^4 \right]\frac{d_{1,1,1}(t,u)}{(t+u)^2} 
\right.\nonumber\\ && \left. 
 +\frac{1}{2}\left[\frac{3}{2}(m_1^2+m_2^2)+\frac{m_1^2 t + 
 m_2^2 u}{t+u}{\bf k}^2 +m_1^2m_2^2(t+u)\right] \frac{d_{1,1,1}(t,u)}{t+u}
\right.\nonumber\\ && \left.
 +\left[\frac{3}{2}-\frac{tu}{t+u}{\bf k}^2\right]\frac{d_{0,0,1}(t,u)}{t+u}\right\}\ .
\label{AppE.6b} 
\end{eqnarray}
\end{subequations}

\subsection{ $1/M$ corrections}              
\label{app:E.b}
The nonadiabatic from the $1/M$-expansion of the energy denominators 
and the pseudo-vector vertex $1/M$-corrections are described in 
Ref.~\cite{Rij91} and used also in Ref.~\cite{RS96b}, section IV.
Below, we give the results for the evaluation of these $1/m$-corrections 
for the 1-pair graph's with the 'derivative' pair-interaction.\\

\noindent {\bf a. Non-adiabatic contributions}: 
For the 1-pair graph's in equation (\ref{eq:3.1}) the non-adiabatic operator is
\begin{eqnarray}
 \tilde{O}^{(1),na}_{\alpha\beta,p}({\bf k}_{1},{\bf k}_2) &\Rightarrow&
 -2\frac{g'_{(\pi\pi)_0}}{m_\pi^3}\left(\frac{f_{NN\pi}}{m_\pi}\right)^2\cdot
 \frac{1}{2M}\cdot
 \left[({\bf k}_1\cdot{\bf k}_2)^2 + \frac{i}{2}\left(
\mbox{\boldmath $\sigma$}_1+ \mbox{\boldmath $\sigma$}_2\right)\cdot
 {\bf k}_1\times{\bf k}_2\ {\bf q}\cdot({\bf k}_1-{\bf k}_2)\right]\cdot 
 \nonumber\\
 && \times \widetilde{\Gamma}^{(S')}_{\pi\pi,p}\ ,
\label{AppE.7} \end{eqnarray}
where 
 $\widetilde{\Gamma}^{(S')}_{\pi\pi,p}= 
 \pm \omega_1\omega_2+{\bf k}_1\cdot{\bf k}_2$ and the $\pm$-sign has been 
explained above. The denominators $D^{(na)}_p(\omega_1,\omega_2)$ have been
given in \cite{RS96b}. Again, we select the terms symmetric in $1 \leftrightarrow 2$
since the asymmetric terms will not contribute, which is easily seen in x-space.
The sum over the graph's $p=a,b,c$ yields
\begin{eqnarray}
 \sum_p \widetilde{\Gamma}^{(S')}_{\pi\pi,p}\ D_p^{na}(\omega_1,\omega_2) &=& 
 \frac{1}{\omega_1\omega_2}\frac{1}{\omega_1+\omega_2}             
 +\frac{1}{\omega_1^2\omega_2^2}
 \left[\frac{1}{\omega_1}+\frac{1}{\omega_2}
 -\frac{1}{\omega_1+\omega_2}\right]\ \left({\bf k}_1\cdot{\bf k}_2\right).
\label{AppE.8} \end{eqnarray}
Using the expressions in this appendix  we obtain  
in p-space the non-adiabatic 'derivative' $(\pi\pi)_0$-exchange potentials
\begin{subequations}
\begin{eqnarray}
&& \Omega_1^{(na)}({\bf k}^2;t,u) = -12
\left(\frac{g'_{(\pi\pi)_0}}{m_\pi^3}\right)
\left(\frac{f_{NN\pi}}{m_\pi}\right)^2\frac{1}{2M}\cdot 
 \left\{\vphantom{\frac{A}{A}}\right. \nonumber\\ & & +     
\left[\frac{15}{4}+\frac{1}{2}\left(\frac{t^2-8tu+u^2}{t+u}\right){\bf k}^2
      +\frac{t^2u^2}{(t+u)^2}{\bf k}^4\right]\frac{d_{1,1,1}(t,u)}{(t+u)^2}\ ,
 \nonumber\\ & & -\left.
\left[\frac{105}{8}+\frac{15}{4}\left(\frac{t^2-5tu+u^2}{t+u}\right){\bf k}^2
      -\frac{3}{2}tu\left(\frac{t^2-5tu+u^2}{(t+u)^2}\right){\bf k}^4
      -\frac{t^3u^3}{(t+u)^3}{\bf k}^6\right]\frac{d_{na}(t,u)}{(t+u)^3}
\right\}\ , \label{AppE.9a}\\
 && \Omega_4^{(na)}({\bf k}^2;t,u) = - 12
\left(\frac{g'_{(\pi\pi)_0}}{m_\pi^3}\right)
\left(\frac{f_{NN\pi}}{m_\pi}\right)^2\frac{1}{2M}\cdot 
\left\{\frac{d_{1,1,1}(t,u)}{t+u}+  
\left[-5+2\frac{tu}{t+u}{\bf k}^2\right]
\frac{d_{na}(t,u)}{(t+u)^2} \right\}\ .
\label{AppE.9b} \end{eqnarray}
\end{subequations}
Here, $d_{\{na\}}(t,u)$ is defined in (\ref{appA.3}).\\

\noindent {\bf b. Pseudo-vector contributions}: 
The pseudovector vertex gives $1/M$-terms as can be seen from 
\begin{equation}
\bar{u}({\bf p}') \Gamma_P^{(1)} u({\bf p}) = -i \frac{f_{NN\pi}}{m_\pi}\left[
 \mbox{\boldmath $\sigma$}\cdot({\bf p}'-{\bf p}) \pm \frac{\omega}{2M}
 \mbox{\boldmath $\sigma$}\cdot({\bf p}'+{\bf p}) \right]\ ,               
\label{AppE.10} \end{equation}
where upper (lower) sign applies for creation (absorption) of the pion at the vertex.
For graph (a) the operator for the nucleon line on the right is readily seen to be
\begin{eqnarray}
&& -\left(\frac{f_P}{m_\pi}\right)^2\frac{1}{2M}\left[ \vphantom{\frac{A}{A}}
 \left(\omega_1{\bf k}_2^2-\omega_2{\bf k}_1^2\right)
 -2{\bf q}\cdot\left(\omega_1{\bf k}_2+\omega_2{\bf k}_1\right)
 +2i\mbox{\boldmath $\sigma$}_2\cdot{\bf q}\times
 \left(\omega_1{\bf k}_2-\omega_2{\bf k}_1\right)\right]
\label{AppE.11} \end{eqnarray}
The same expression for gragh (b) is obviously obtained from (\ref{AppE.11}) by 
making the
the substitution $\omega_1 \rightarrow -\omega_1$, and for graph (c) the 
substitution $\omega_{1,2} \rightarrow -\omega_{1,2}$.                       
The mirror graphs are included by making the 
replacement $\mbox{\boldmath $\sigma$}_2 \rightarrow
\mbox{\boldmath $(\sigma$}_1 + \mbox{\boldmath $\sigma$}_2)/2$. 
Combining all this with the adiabatic denominators 
$D^{1}_i(\omega_1,\omega_2)$ 
\begin{equation}
D^{1}_a(\omega_1,\omega_2)= 
\frac{1}{2\omega_1\omega_2^2(\omega_1+\omega_2)}\ \ ,\ \ 
D^{1}_b(\omega_1,\omega_2)= 
\frac{1}{2\omega_1^2\omega_2^2}\ ,              
\label{AppE.12} \end{equation}
and $D^{1}_c(\omega_1,\omega_2)=D^{1}_a(\omega_2,\omega_1)$.
Summing over the 1-pair graphs gives
\begin{eqnarray}
 && \sum_p \tilde{O}^{(1),na}_{\alpha\beta,p}({\bf k}_{1},{\bf k}_2) 
 D^{1}_p(\omega_1,\omega_2)= 
 -\frac{g'_{(\pi\pi)_0}}{m_\pi^3}\left(\frac{f_{NN\pi}}{m_\pi}\right)^2\cdot
\frac{1}{M}\cdot \frac{1}{\omega_1\omega_2\left(\omega_1+\omega_2\right)}
\cdot\nonumber\\  && \times
 \left[ \left\{\frac{1}{2}(m_1^2+m_2^2)(\omega_1^2+\omega_2^2)
 -2\omega_1^2\omega_2^2\right\}
 -({\bf k}_1^2+{\bf k}_2^2)({\bf k}_1\cdot{\bf k}_2)
 \vphantom{\frac{A}{A}}\right. \nonumber\\ 
 && \left. \vphantom{\frac{A}{A}} 
 -i\left(\mbox{\boldmath $\sigma$}_1 + \mbox{\boldmath $\sigma$}_2 \right)
 \cdot{\bf q}\times{\bf k}
 \left(\omega_1^2+\omega_2^2+{\bf k}_1\cdot{\bf k}_2\right)
\right]\ .
\label{AppE.13} \end{eqnarray}
Using the expressions in this appendix  we obtain  
in p-space the pseudo-vector vertex $1/M$-corrections to the 
'derivative' $(\pi\pi)_0$-exchange potentials
\begin{subequations}
\begin{eqnarray}
\Omega_1^{(1),pv}({\bf k}^2;t,u) &=& -6 
\left(\frac{g'_{(\pi\pi)_0}}{m_\pi^3}\right)
\left(\frac{f_{NN\pi}}{m_\pi}\right)^2\frac{1}{2M}\cdot d_{1,1,1}(t,u)\cdot 
\left\{\vphantom{\frac{A}{A}} \left(m_1^2-m_2^2\right)^2 \right.
 \nonumber\\ & &        
-\left[3(m_1^2+m_2^2)+\frac{m_1^2(3t^2-u^2)+m_2^2(3u^2-t^2)}{t+u}{\bf k}^2
 \right]\frac{1}{t+u} \nonumber\\ && \left. -
 \left[\left(\frac{t^2+2tu+u^2}{t+u}\right){\bf k}^2 +          
      2tu\left(\frac{t^2+2tu+u^2}{(t+u)^2}\right){\bf k}^4\right]
 \frac{1}{(t+u)^2} \right\}\ ,
\label{AppE.14a}\\              
\Omega_4^{(1),pv}({\bf k}^2;t,u) &=& -24
\left(\frac{g'_{(\pi\pi)_0}}{m_\pi^3}\right)
\left(\frac{f_{NN\pi}}{m_\pi}\right)^2\frac{1}{2M}\cdot d_{1,1,1}(t,u)\cdot 
 \nonumber\\ & & \times
 \left\{ \left(m_1^2+m_2^2\right) + 
 \left[\frac{3}{2}+\left(\frac{t^2+tu+u^2}{t+u}\right){\bf k}^2\right]
 \frac{1}{t+u}\right\}\ .
\label{AppE.14b} \end{eqnarray}
\end{subequations}

\end{widetext}

\section{Pair Couplings and $SU_f(3)$-symmetry}
\label{app:F.a}     
Below, $\sigma, {\bf a}_0, {\bf A}_1, \ldots $ are short-hands for 
respectively  the nucleon densities $\bar{\psi} \psi$,
$\bar{\psi}\mbox{\boldmath $\tau$}\psi$,
$\bar{\psi}\gamma_5\gamma_\mu\mbox{\boldmath $\tau$}\psi, \ldots $.

The $SU_f(3)$ octet and singlet mesons, denoted by the subscript $8$
respectively $1$, are in terms of the physical ones defined as follows:
\begin{enumerate}
\item[(i)] \underline{Pseudo-scalar-mesons}:
\begin{eqnarray*}
   \eta_1 &=& \cos\theta_{pv} \eta' - \sin\theta_{pv} \eta \\       
   \eta_8 &=& \sin\theta_{pv} \eta' + \cos\theta_{pv} \eta              
\end{eqnarray*}
Here, $\eta'$ and $\eta$ are the physical pseudo-scalar mesons 
 $\eta(957)$ respectively $\eta(548)$.
\item[(ii)] \underline{Vector-mesons}:         
\begin{eqnarray*}
   \phi_1 &=& \cos\theta_{v} \omega  - \sin\theta_{v} \phi \\       
   \phi_8 &=& \sin\theta_{v} \omega + \cos\theta_{v} \phi              
\end{eqnarray*}
Here, $\phi$ and $\omega$ are the physical vector mesons 
 $\phi(1019)$ respectively $\omega(783)$.
\end{enumerate}
Then, one has the following $SU(3)$-invariant pair-interaction 
Hamiltonians:\\
 1.\ $SU(3)$-singlet couplings $S^\alpha_\beta = 
\delta^\alpha_\beta \sigma/\sqrt{3}$:
\begin{eqnarray*}
 {\cal H}_{S_1PP} &=& \frac{g_{S_1PP}}{\sqrt{3}}\left\{
\mbox{\boldmath $\pi$}\cdot\mbox{\boldmath $\pi$} + 
  2 K^\dagger K + \eta_8\eta_8\right\}\cdot \sigma
\end{eqnarray*}
 2.\ $SU(3)$-octet symmetric couplings I, 
 $S^\alpha_\beta = (S_8)^\alpha_\beta \Rightarrow (1/4) Tr\{ S[P,P]_+\}$:
\begin{eqnarray*}
 {\cal H}_{S_8PP} &=& 
\frac{g_{S_8PP}}{\sqrt{6}}\left\{\vphantom{\frac{A}{A}}\right.
 ({\bf a}_0\cdot\mbox{\boldmath $\pi$})\eta_8 + 
 \frac{\sqrt{3}}{2}{\bf a}_0\cdot(K^\dagger \mbox{\boldmath $\tau$}K) 
 \nonumber \\
 && +\frac{\sqrt{3}}{2}
 \left\{(K_0^\dagger\mbox{\boldmath $\tau$}K)\cdot\mbox{\boldmath $\pi$}+ 
 h.c.\right\} \nonumber\\ &&
 -\frac{1}{2}\left\{(K_0^\dagger K)\eta_8 + h.c. \right\} 
 \nonumber \\ && 
 + \frac{1}{2}f_0\left(\mbox{\boldmath $\pi$}\cdot\mbox{\boldmath $\pi$} 
 - K^\dagger K -\eta_8\eta_8\right) 
 \left.\vphantom{\frac{A}{A}}\right\}
\end{eqnarray*}
 3.\ $SU(3)$-octet symmetric couplings II, 
 $S^\alpha_\beta = (B_8)^\alpha_\beta \Rightarrow (1/4) Tr\{ B^\mu [V_\mu, P]_+\}$:
\begin{eqnarray*}
 {\cal H}_{B_8VP} &=&
 \frac{g_{B_8VP}}{\sqrt{6}}\left\{\vphantom{\frac{A}{A}}\right. 
 \frac{1}{2}\left[\left({\bf B}_1^\mu\cdot\mbox{\boldmath $\rho$}_\mu\right) \eta_8 +
 \left({\bf B}_1^\mu\cdot\mbox{\boldmath $\pi$}_\mu\right) \phi_8 \right] 
  \nonumber \\ &&
 +\frac{\sqrt{3}}{4}\left[{\bf B}_1\cdot(K^{*\dagger}\mbox{\boldmath $\tau$}K)
 + h.c. \right] 
  \nonumber \\ &&
 +\frac{\sqrt{3}}{4}\left[(K_1^\dagger\mbox{\boldmath $\tau$} K^*)\cdot
 \mbox{\boldmath $\pi$}
 +(K_1^\dagger\mbox{\boldmath $\tau$} K)\cdot\mbox{\boldmath $\rho$} + h.c. \right] 
  \nonumber \\ &&
 -\frac{1}{4}\left[(K_1^\dagger\cdot K^*) \eta_8 + (K_1^\dagger\cdot K) \phi_8 
 + h.c. \right] 
  \nonumber \\ &&
 +\frac{1}{2}H^0\left[\mbox{\boldmath $\rho$}\cdot\mbox{\boldmath $\pi$} 
 -\frac{1}{2}\left(K^{*\dagger}\cdot K+ K^\dagger\cdot K^* \right)
 -\phi_8\eta_8 \right]
 \left.\vphantom{\frac{A}{A}}\right\}                            
\end{eqnarray*}
 4.\ $SU(3)$-octet a-symmetric couplings I, 
 $A^\alpha_\beta = (V_8)^\alpha_\beta \Rightarrow 
 (-i/\sqrt{2}) Tr\{ V^\mu [P,\partial_\mu P]_-\}$:
\begin{eqnarray*}
 {\cal H}_{V_8PP} &=& g_{A_8PP}\left\{\vphantom{\frac{A}{A}}\right.
 \frac{1}{2}\mbox{\boldmath $\rho$}_\mu\cdot\mbox{\boldmath $\pi$}\times
 \stackrel{\leftrightarrow}{\partial^\mu}\!\!
 \mbox{\boldmath $\pi$}+\frac{i}{2}\mbox{\boldmath $\rho$}_\mu\cdot(K^\dagger
 \mbox{\boldmath $\tau$}\!\!\stackrel{\leftrightarrow}{\partial^\mu}\!\! K) 
 \nonumber \\
 && +\frac{i}{2}\left(\vphantom{\frac{A}{A}} K^{* \dagger}_\mu \mbox{\boldmath $\tau$}
(K\!\!\stackrel{\leftrightarrow}{\partial^\mu}\!\!\mbox{\boldmath $\pi$}) 
 - h.c. \right)
 +i\frac{\sqrt{3}}{2}\left(\vphantom{\frac{A}{A}} K^{* \dagger}_\mu\cdot
 \right.\nonumber\\  && \left. (K\cdot\stackrel{\leftrightarrow}
{\partial^\mu}\!\! \eta_8) - h.c. \vphantom{\frac{A}{A}}\right) 
 +\frac{i}{2}\sqrt{3} \phi_\mu (K^\dagger\stackrel{\leftrightarrow}{\partial^\mu}\!\! K)
 \left.\vphantom{\frac{A}{A}}\right\}
\end{eqnarray*}
 5.\ $SU(3)$-octet a-symmetric couplings II, 
 $A^\alpha_\beta = (A_8)^\alpha_\beta \Rightarrow 
 (-i/\sqrt{2}) Tr\{ A^\mu [P,V_\mu]_-\}$:
\begin{eqnarray*}
 {\cal H}_{A_8VP} &=& g_{A_8VP}\left\{\vphantom{\frac{A}{A}}\right.
 {\bf A}_1\cdot\mbox{\boldmath $\pi$}\times\mbox{\boldmath $\rho$}
 \nonumber\\ &&
 +\frac{i}{2}{\bf A}_1\cdot\left[(K^\dagger\mbox{\boldmath $\tau$} K^*)  
  -(K^{*\dagger}\mbox{\boldmath $\tau$} K)\right] \nonumber \\
 &&
 -\frac{i}{2}\left(\left[(K^\dagger\mbox{\boldmath $\tau$}K_A)\cdot
 \mbox{\boldmath $\rho$}
 + (K_{A}^\dagger\mbox{\boldmath $\tau$}K^*)\cdot\mbox{\boldmath $\pi$}
 \right] - h.c.\right) \nonumber \\
 &&
 -i\frac{\sqrt{3}}{2}\left(\left[(K^\dagger\cdot K_A)\phi_8
 +(K_A^\dagger\cdot K^*)\eta_8\right] - h.c. \right) \nonumber \\
 &&
 +\frac{i}{2}\sqrt{3} f_1 \left[K^\dagger\cdot K^*-K^{*\dagger}\cdot K\right]  
 \left.\vphantom{\frac{A}{A}}\right\}
\end{eqnarray*}
The relation with the pair-couplings of Appendix~\ref{app:AA} is 
 $g_{S_1PP}/\sqrt{3}= g_{(\pi\pi)_0}/m_\pi$, 
$g_{A_8VP}= g_{(\pi\rho)_1}/m_\pi$ etc.

\begin{widetext}
 
\begin{table*}[h]
\caption{The one-pair isospin factors $C^{(1)}(\alpha\beta)$ and
         momentum operators $\tilde{O}^{(1)}_{\alpha\beta,p}({\bf k}_{1},
         {\bf k}_{2})$. The index $p$ labels the type of denominators.
         Note that $\kappa_{1}=(f/g)_{(\pi\pi)_{1}}$.}
\begin{ruledtabular}
\begin{tabular}{lll}
 &&\\
 $(\alpha\beta)$ & $C^{(1)}(\alpha\beta)$ &
 $ O_{\alpha\beta,p}^{(1)}({\bf k}_{1},{\bf k}_{2})$\\
 &&\\
\colrule
 &&\\
 $(\pi\pi)_{0}$   & $6$                                          &
      $-{\bf k}_{1}\!\cdot{\bf k}_{2}
        +{\textstyle\frac{i}{2}}(\mbox{\boldmath$\sigma$}_{1} + \mbox{\boldmath$\sigma$}_{2} )
         \!\cdot\!({\bf k}_{1}\times{\bf k}_{2}) $             \\[0.2cm]
 $(\sigma\sigma)$ & $2$                                &  $1$  \\[0.2cm]
 $(\pi\eta)$      & $\mbox{\boldmath$\tau$}_{1}\!\cdot\!\mbox{\boldmath$\tau$}_{2}$    &
      $-2{\bf k}_{1}\!\cdot{\bf k}_{2}$                        \\[0.2cm]
 $(\pi\eta')$     & $\mbox{\boldmath$\tau$}_{1}\!\cdot\!\mbox{\boldmath$\tau$}_{2}$    &
      $-2{\bf k}_{1}\!\cdot{\bf k}_{2}$                        \\[0.2cm]
 $(\pi\pi)_{1}$   & $2i\mbox{\boldmath$\tau$}_{1}\!\cdot\!\mbox{\boldmath$\tau$}_{2}$ &
      $i\left[{\bf k}_{1}\!\cdot\!{\bf k}_{2}
         -{\textstyle\frac{i}{2}}(\mbox{\boldmath$\sigma$}_{1}+\mbox{\boldmath$\sigma$}_{2})
         \!\cdot\!({\bf k}_{1}\times{\bf k}_{2})\right] $     \\[0.2cm]
  & & $ \hspace{0ex} + {\displaystyle\frac{i}{M}} \left[ (1+\kappa_{1})
       \mbox{\boldmath$\sigma$}_{1}\!\cdot\!({\bf k}_{1}\times{\bf k}_{2})
       \mbox{\boldmath$\sigma$}_{2}\!\cdot\!({\bf k}_{1}\times{\bf k}_{2})\right.$ \\[0.2cm]
  & & $\left. +{\textstyle\frac{i}{2}}(\mbox{\boldmath$\sigma$}_{1}+\mbox{\boldmath$\sigma$}_{2})
         \!\cdot\!({\bf k}_{1}\times{\bf k}_{2})\,{\bf q}\!\cdot\!
                  ({\bf k}_{1}-{\bf k}_{2})\right]$            \\[0.2cm]
 $(\pi\rho)_{1}$  & $-2i\mbox{\boldmath$\tau$}_{1}\!\cdot\!\mbox{\boldmath$\tau$}_{2}$ &
      ${\displaystyle\frac{i}{M}}
       \left[\mbox{\boldmath$\sigma$}_{1}\!\cdot\!{\bf k}_{1}
             \mbox{\boldmath$\sigma$}_{2}\!\cdot\!{\bf k}_{1}
         +{\textstyle\frac{1}{2}}(1+\kappa_{\rho})
          \left(\mbox{\boldmath$\sigma$}_{1}\!\cdot\!{\bf k}_{1}
           \mbox{\boldmath$\sigma$}_{2}\!\cdot\!{\bf k}_{2}  \right.\right.$  \\[0.2cm]
  & & $  \left.\left. + \mbox{\boldmath$\sigma$}_{1}\!\cdot\!{\bf k}_{2}                         
           \mbox{\boldmath$\sigma$}_{2}\!\cdot\!{\bf k}_{1} -
          2\mbox{\boldmath$\sigma$}_{1}\!\cdot\!\mbox{\boldmath$\sigma$}_{2}
          {\bf k}_{1}\!\cdot\!{\bf k}_{2}\right) \right]$           \\[0.2cm]
 $(\pi\sigma)$    & $\mbox{\boldmath$\tau$}_{1}\!\cdot\!\mbox{\boldmath$\tau$}_{2}$    &
      $\left[\mbox{\boldmath$\sigma$}_{1}\!\cdot\!{\bf k}_{1}
             \mbox{\boldmath$\sigma$}_{2}\!\cdot\!{\bf k}_{2} +
             \mbox{\boldmath$\sigma$}_{1}\!\cdot\!{\bf k}_{2}
             \mbox{\boldmath$\sigma$}_{2}\!\cdot\!{\bf k}_{1} -
            2\mbox{\boldmath$\sigma$}_{1}\!\cdot\!{\bf k}_{1}
             \mbox{\boldmath$\sigma$}_{2}\!\cdot\!{\bf k}_{1} \right] $  \\[0.2cm]
 $(\pi P)$        & $\mbox{\boldmath$\tau$}_{1}\!\cdot\!\mbox{\boldmath$\tau$}_{2}$    &
      $\left[\mbox{\boldmath$\sigma$}_{1}\!\cdot\!{\bf k}_{1}
             \mbox{\boldmath$\sigma$}_{2}\!\cdot\!{\bf k}_{2} +
             \mbox{\boldmath$\sigma$}_{1}\!\cdot\!{\bf k}_{2}
             \mbox{\boldmath$\sigma$}_{2}\!\cdot\!{\bf k}_{1} -
            2\mbox{\boldmath$\sigma$}_{1}\!\cdot\!{\bf k}_{1}
             \mbox{\boldmath$\sigma$}_{2}\!\cdot\!{\bf k}_{1} \right] $  \\[0.2cm]
 $(\pi\rho)_{0}$  & $3$                                          &
      $\left[\mbox{\boldmath$\sigma$}_{1}\!\cdot\!{\bf k}_{1}
             \mbox{\boldmath$\sigma$}_{2}\!\cdot\!{\bf k}_{2} +
             \mbox{\boldmath$\sigma$}_{1}\!\cdot\!{\bf k}_{2}
             \mbox{\boldmath$\sigma$}_{2}\!\cdot\!{\bf k}_{1} +
            2\mbox{\boldmath$\sigma$}_{1}\!\cdot\!{\bf k}_{1}
             \mbox{\boldmath$\sigma$}_{2}\!\cdot\!{\bf k}_{1} \right] $  \\[0.2cm]
 $(\pi\omega)$    & $\mbox{\boldmath$\tau$}_{1}\!\cdot\!\mbox{\boldmath$\tau$}_{2}$    &
      $\left[\mbox{\boldmath$\sigma$}_{1}\!\cdot\!{\bf k}_{1}
             \mbox{\boldmath$\sigma$}_{2}\!\cdot\!{\bf k}_{2} +
             \mbox{\boldmath$\sigma$}_{1}\!\cdot\!{\bf k}_{2}
             \mbox{\boldmath$\sigma$}_{2}\!\cdot\!{\bf k}_{1} +
            2\mbox{\boldmath$\sigma$}_{1}\!\cdot\!{\bf k}_{1}
             \mbox{\boldmath$\sigma$}_{2}\!\cdot\!{\bf k}_{1} \right] $
\end{tabular}
\end{ruledtabular}
\label{O1pair}
\end{table*} 

\begin{table*}[h]
\caption{The one-pair denominators $D^{(1)}_p(\omega_1,\omega_2)$}             \begin{ruledtabular}
\begin{tabular}{llll}
 &&&\\
 $(\alpha\beta)$ & $D^{(1)}_{ad}(\omega_1,\omega_2)$&
 $ D^{(1)}_{na}(\omega_1,\omega_2)$& $ D^{(1)}_{pv}(\omega_1,\omega_2)$ \\                  
 &&&\\
\colrule
 &&&\\
 $(\pi\pi)_{0}$   & $ \displaystyle{\frac{1}{\omega_1^2\omega_2^2}}$ &
 $\displaystyle{\frac{1}{\omega_1^2\omega_2^2}\left[\frac{1}{\omega_1}+
 \frac{1}{\omega_2}-\frac{1}{\omega_1+\omega_2}\right]}$ &
 $\displaystyle{\frac{1}{\omega_1\omega_2(\omega_1+\omega_2)}}$         \\[0.2cm]
 &&&\\
 $(\pi\pi)_{1}$   & $\displaystyle{\frac{2}{\omega_1\omega_2(\omega_1+\omega_2)}\ \ , \ \
 \frac{1}{\omega_1^2\omega_2^2}}$&
 $\displaystyle{\frac{2}{\omega_1^2\omega_2^2}}$ & 
 $\displaystyle{\frac{1}{\omega_1^2}}$\ ,\ $\displaystyle{\frac{1}{\omega_2^2}}$ \\[0.2cm]
 &&&\\
 $(\pi\sigma)$  & $ \displaystyle{\frac{1}{\omega_1^2\omega_2^2}}$ &
 $\displaystyle{\frac{1}{\omega_1^2\omega_2^2}\left[\frac{1}{\omega_1}+
 \frac{1}{\omega_2}-\frac{1}{\omega_1+\omega_2}\right]}$ &
 $\displaystyle{\frac{1}{\omega_1\omega_2(\omega_1+\omega_2)}}$           \\[0.2cm]
\end{tabular}
\end{ruledtabular}
\label{D1pair}
\end{table*}
\begin{table*}[h]
\caption{The two-pair isospin factors $C^{(2)}(\alpha\beta)$ and
         momentum operators $\tilde{O}^{(2)}_{\alpha\beta,p}({\bf k}_{1},
         {\bf k}_{2})$, and denominators $D^{(2)}_p(\omega_1,\omega_2)$.}  
\begin{ruledtabular}
\begin{tabular}{lllc}
 &&&\\
 $ (\alpha\beta)$ & $C^{(2)}(\alpha\beta)$ &
 $ \tilde{O}_{\alpha\beta,p}^{(2)}({\bf k}_{1},{\bf k}_{2})$ &
 $ D^{(2)}_p(\omega_1,\omega_2)$  \\
 &&&\\
\colrule
 &&&\\
 $(\pi\pi)_{0}$   & $6$ & $1$ & $-\displaystyle{\frac{1}{2\omega_1\omega_2}} 
 \displaystyle{\frac{1}{\omega_1+\omega_2}}$  \\[0.2cm]
 $(\sigma\sigma)$ & $2$ & $1$ & ,, \\[0.2cm]
 $(\pi\eta)$      & $\mbox{\boldmath$\tau$}_{1}\!\cdot\!\mbox{\boldmath$\tau$}_{2}$    &
      $1$         & ,,                                          \\[0.2cm]
 $(\pi\eta')$     & $\mbox{\boldmath$\tau$}_{1}\!\cdot\!\mbox{\boldmath$\tau$}_{2}$    &
      $1$         & ,,                                          \\[0.2cm]
 $(\pi\pi)_{1}$   & $\mbox{\boldmath$\tau$}_{1}\!\cdot\!\mbox{\boldmath$\tau$}_{2}$    &
      $1$         & $ -\displaystyle{\frac{1}{2}\left[\frac{1}{\omega_1} 
 +\frac{1}{\omega_2}-\frac{4}{\omega_1+\omega_2}\right]}$         \\[0.2cm]
 $(\pi\rho)_{1}$  & $2\mbox{\boldmath$\tau$}_{1}\!\cdot\!\mbox{\boldmath$\tau$}_{2}$   &
      $\mbox{\boldmath$\sigma$}_{1}\!\cdot\!\mbox{\boldmath$\sigma$}_{2}$ 
 & $-\displaystyle{\frac{1}{2\omega_1\omega_2}}  
 \displaystyle{\frac{1}{\omega_1+\omega_2}}$                     \\[0.2cm]
 $(\pi\sigma)$    & $\mbox{\boldmath$\tau$}_{1}\!\cdot\!\mbox{\boldmath$\tau$}_{2}$    &
      $\mbox{\boldmath$\sigma$}_{1}\!\cdot\!({\bf k}_{1}-{\bf k}_{2})
       \mbox{\boldmath$\sigma$}_{2}\!\cdot\!({\bf k}_{1}-{\bf k}_{2})$ & ,, \\[0.2cm]
 $(\pi P)$        & $\mbox{\boldmath$\tau$}_{1}\!\cdot\!\mbox{\boldmath$\tau$}_{2}$    &
      $\mbox{\boldmath$\sigma$}_{1}\!\cdot\!({\bf k}_{1}-{\bf k}_{2})
       \mbox{\boldmath$\sigma$}_{2}\!\cdot\!({\bf k}_{1}-{\bf k}_{2})$ & ,, \\[0.2cm]
 $(\pi\rho)_{0}$  & $3$                                          &
      $\mbox{\boldmath$\sigma$}_{1}\!\cdot\!\mbox{\boldmath$\sigma$}_{2}$              &
      $\displaystyle{-\frac{1}{2}\left(\frac{1}{\omega_1}+\frac{1}{\omega_2}\right)}$\\[0.2cm]
                  &                                              &
          $\displaystyle{-\mbox{\boldmath$\sigma$}_{1}\!\cdot\!({\bf k}_{1}+{\bf k}_{2})
          \mbox{\boldmath$\sigma$}_{2}\!\cdot\!({\bf k}_{1}+{\bf k}_{2})}$ & 
 $\displaystyle{-\frac{1}{2\omega_1\omega_2}\frac{1}{\omega_1+\omega_2}}$ \\[0.2cm]
 $(\pi\omega)$    & $\mbox{\boldmath$\tau$}_{1}\!\cdot\!\mbox{\boldmath$\tau$}_{2}$    &
      $\mbox{\boldmath$\sigma$}_{1}\!\cdot\!\mbox{\boldmath$\sigma$}_{2}$              &
      $\displaystyle{-\frac{1}{2}\left(\frac{1}{\omega_1}+\frac{1}{\omega_2}\right)}$\\[0.2cm]
                  &                                              &
          $\displaystyle{-\mbox{\boldmath$\sigma$}_{1}\!\cdot\!({\bf k}_{1}+{\bf k}_{2})
          \mbox{\boldmath$\sigma$}_{2}\!\cdot\!({\bf k}_{1}+{\bf k}_{2})}$ & 
 $\displaystyle{-\frac{1}{2\omega_1\omega_2}\frac{1}{\omega_1+\omega_2}}$ \\
\end{tabular}
\end{ruledtabular}
\label{O2pair} 
\end{table*}
\begin{table*}[h]
\caption{The $d_p(t,u)$-functions cooresponding to the denominators 
         $D_p(\omega_1,\omega_2)\ ,\ p=0,1,2,3,4,5$.}
\begin{ruledtabular}
\begin{tabular}{ccl|ccl}
 &&&\\
 \multicolumn{3}{c|}{$ D_{\{p\}}(\omega_1,\omega_2)$} & 
 \multicolumn{3}{c}{$ d_{\{p\}}(t,u))$} \\
 &&&\\
\colrule
 &&&\\
 $\displaystyle{D_{2,2,0}}$ &=& $\displaystyle{\frac{1}{\omega_1^2\omega_2^2}}$ & 
 $\displaystyle{d_{2,2,0}}$ &=& $\displaystyle{ 1 }$\\[0.2cm]
 $\displaystyle{D_{2,0,0}}$ &=& $\displaystyle{\frac{1}{\omega_1^2}}$ & 
 $\displaystyle{d_{2,0,0}}$ &=& $\displaystyle{ \delta(u-u_0)}$\\[0.2cm]
 $\displaystyle{D_{0,2,0}}$ &=& $\displaystyle{\frac{1}{\omega_2^2}}$ & 
 $\displaystyle{d_{0,2,0}}$ &=& $\displaystyle{ \delta(t-t_0)}$\\[0.2cm]
 $\displaystyle{D_{1,0,0}}$ &=& $\displaystyle{\frac{1}{\omega_1}}$ & 
 $\displaystyle{d_{1,0,0}}$ &=& $\displaystyle{\frac{1}{\sqrt{\pi}} t^{-1/2}\
 \delta(u-u_0)}$\\[0.2cm]
 $\displaystyle{D_{0,1,0}}$ &=& $\displaystyle{\frac{1}{\omega_2}}$ & 
 $\displaystyle{d_{0,1,0}}$ &=& $\displaystyle{\frac{1}{\sqrt{\pi}} u^{-1/2}\
 \delta(t-t_0)}$\\[0.2cm]
 $\displaystyle{D_{0,0,1}}$ &=& $\displaystyle{\frac{1}{\omega_1+\omega_2}}$ & 
 $\displaystyle{d_{0,0,1}}$ &=& $\displaystyle{\frac{1}{2\sqrt{\pi}} (t+u)^{-3/2}}$\\[0.2cm]
 $\displaystyle{D_{1,1,1}}$ &=& $\displaystyle{\frac{1}{\omega_1\omega_2} 
 \frac{1}{\omega_1+\omega_2}}$ & $\displaystyle{d_{1,1,1}}$ 
 &=& $\displaystyle{\frac{1}{\sqrt{\pi}} (t+u)^{-1/2}}$\\[0.2cm]
\end{tabular}
\end{ruledtabular}
\label{dtus}   
\end{table*}
\begin{table*}[h]
\caption{Coefficients \protect$\Upsilon^{(ad,na,pv)}_{j,k}$ for the 
         \protect$(\pi\pi)_0$ contributions.}                          
\begin{ruledtabular}
\begin{tabular}{c|lll}  
 &&&\\
    & $\Upsilon_{0}(//)(t,u)$ & $\Upsilon_1(//)(t,u)$ & $\Upsilon_2(//)(t,u)$ \\            
&&&\\
\colrule
&&&\\
  & \multicolumn{3}{c}{ 1-pair exchange} \\
&&&\\
\colrule
&&&\\
 $\Omega^{(1),ad}_1$ & $\displaystyle{+\frac{3}{2}\frac{1}{t+u}}$ & 
 $\displaystyle{-\frac{tu}{(t+u)^2} }$ & --- \\[3mm]        
&&&\\
 $\Omega^{(1),na}_1$ & $\displaystyle{-\frac{1}{\sqrt{\pi}}\frac{15}{4}
 \frac{\sqrt{t+u}}{(t+u)^2}\cdot\frac{1}{M}}$ & 
 $\displaystyle{-\frac{1}{2\sqrt{\pi}}\left(\frac{t^2-8tu+u^2}
 {t+u}\right)\frac{\sqrt{t+u}}{(t+u)^2}\cdot\frac{1}{M}}$ & 
 $\displaystyle{-\frac{1}{\sqrt{\pi}}
 \left(\frac{t^2u^2}{(t+u)^2}\right)\frac{\sqrt{t+u}}{(t+u)^2}\cdot\frac{1}{M}}$ \\[3mm]
 $\Omega^{(1),na}_4$ & $\displaystyle{-\frac{1}{\sqrt{\pi}}
 \frac{\sqrt{t+u}}{t+u}\cdot\frac{1}{M}}$ & --- & --- \\[3mm]                                            
&&&\\
 $\Omega^{(1),pv}_1$ & $\displaystyle{\frac{3}{2\sqrt{\pi}}
 \frac{1}{(t+u)^{3/2}}\cdot\frac{1}{M}}$ & 
 $\displaystyle{\frac{1}{2\sqrt{\pi}}\left(\frac{t^2+u^2}
 {t+u}\right)\frac{1}{(t+u)^{3/2}}\cdot\frac{1}{M}}$ & --- \\[3mm]                         
 $\Omega^{(1),pv}_4$ & $\displaystyle{+\frac{3}{\sqrt{\pi}}
 \frac{1}{(t+u)^{1/2}}\cdot\frac{1}{M} }$ & --- & --- \\[3mm]                                        
&&&\\
\colrule
&&&\\
  & \multicolumn{3}{c}{ 2-pair exchange} \\
&&&\\
\colrule
&&&\\
 $\Omega^{(2),ad}_1$ & $\displaystyle{-\frac{1}{2\sqrt{\pi}}
 \frac{1}{(t+u)^{1/2}}}$ & --- & --- \\[3mm]                                        
\end{tabular}
\end{ruledtabular}
\label{table1}
\end{table*}
\begin{table*}[h]
\caption{Coefficients \protect$\Upsilon^{(ad,na,pv)}_{j,k}$ for the 
         \protect$(\pi\pi)_1$ contributions.}                          
\begin{ruledtabular}
\begin{tabular}{c|lll}  
 &&&\\
    & $\Upsilon_{0}(//)(t,u)$ & $\Upsilon_1(//)(t,u)$ & $\Upsilon_2(//)(t,u)$ \\            
&&&\\
\colrule
&&&\\
  & \multicolumn{3}{c}{ 1-pair exchange} \\
&&&\\
\colrule
&&&\\
 $\Omega^{(1),ad}_1$ & $\displaystyle{+\frac{3}{\sqrt{\pi}}\frac{1}{(t+u)^{3/2} }}$ & 
 $\displaystyle{-\frac{2}{\sqrt{\pi}}\frac{tu}{(t+u)^{5/2}} }$ & --- \\[3mm]        
 $\Omega^{(1),ad}_2$ & --- & $\displaystyle{-\frac{1}{3\sqrt{\pi}}
 \frac{(1+\kappa_1)}{M}\frac{1}{t+u} }$  & --- \\[3mm]        
 $\Omega^{(1),ad}_3$ & $\displaystyle{+\frac{1}{2\sqrt{\pi}}
 \frac{(1+\kappa_1)}{M}\frac{1}{t+u} }$  & --- & --- \\[3mm]        
 $\Omega^{(1),ad}_4$ & $\displaystyle{-\frac{1}{\sqrt{\pi}}
 \frac{1}{M}\frac{1}{t+u} }$  & --- & --- \\[3mm]        
&&&\\
 $\Omega^{(1),na}_1$ & $\displaystyle{-\frac{15}{4} \frac{1}{(t+u)^2}\cdot\frac{1}{M}}$ & 
 $\displaystyle{-\frac{1}{2}\left(\frac{t^2-8tu+u^2}{(t+u)^3}\right)\cdot\frac{1}{M}}$ & 
 $\displaystyle{-\frac{t^2u^2}{(t+u)^4}\cdot\frac{1}{M} }$ \\[3mm]   
 $\Omega^{(1),na}_4$ & $\displaystyle{-\frac{1}{t+u}\cdot\frac{1}{M} }$ & 
 --- & ---  \\[3mm]   
&&&\\
 $\Omega^{(1),pv}_1$ & $\displaystyle{+\frac{3}{4}
 \frac{\delta(t-t_0)+\delta(u-u_0)}{t+u}\cdot\frac{1}{M} }$ & 
 $\displaystyle{+\frac{1}{2}\frac{t^2\delta(t-t_0)+u^2\delta(u-u_0)}          
 {(t+u)^2}\cdot\frac{1}{M} }$ & --- \\[3mm]
 $\Omega^{(4),pv}_1$ & $\displaystyle{+
 \frac{t\ \delta(t-t_0)+u\ \delta(u-u_0)}{t+u}\cdot\frac{1}{M} }$ & 
   --- & --- \\[3mm]
&&&\\
\colrule
&&&\\
  & \multicolumn{3}{c}{ 2-pair exchange} \\
&&&\\
\colrule
&&&\\
 $\Omega^{(2),ad}_1$ & \multicolumn{2}{l}{
 $\displaystyle{-\frac{1}{2\sqrt{\pi}}
 \left[\frac{\delta(u-u_0)}{\sqrt{t}}+\frac{\delta(t-t_0)}{\sqrt{u}} -
 \frac{2}{(t+u)^{3/2}}\right] }$ } &  \ \ --- \\[3mm]
\end{tabular}
\end{ruledtabular}
\label{table2}
\end{table*}
\begin{table*}[h]
\caption{Coefficients \protect$\Upsilon^{(ad,na,pv)}_{j,k}$ for the 
         \protect$(\pi\rho)_1$ contributions.}                          
\begin{ruledtabular}
\begin{tabular}{c|ccc}
 &&&\\
    & $\Upsilon_{0}(//)(t,u)$ & $\Upsilon_1(//)(t,u)$ & $\Upsilon_2(//)(t,u)$ \\            
&&&\\
\colrule
&&&\\
  & \multicolumn{3}{c}{ 1-pair exchange} \\
&&&\\
\colrule
&&&\\
 $\Omega^{(1),ad}_2$ & $\displaystyle{+\frac{1}{M}\left(\frac{3}{2}+\kappa_\rho\right)
 \frac{1}{t+u} }$ &
 $\displaystyle{\frac{1}{3M}\left(\frac{u^2}{(t+u)^2}-2(1+\kappa_\rho) 
 \frac{tu}{(t+u)^2}\right) }$ & --- \\[3mm]
 $\Omega^{(1),ad}_3$ & $\displaystyle{+\frac{1}{M}
 \frac{u^2+tu \left(1+\kappa_\rho\right)}{(t+u)^2} }$ & --- & --- \\[3mm]
&&&\\
\colrule
&&&\\
  & \multicolumn{3}{c}{ 2-pair exchange} \\
&&&\\
\colrule
&&&\\
 $\Omega^{(2),ad}_2$ & $\displaystyle{-\frac{1}{2\sqrt{\pi}}
 \frac{1}{(t+u)^{1/2}}}$ & --- & ---- \\[3mm]
\end{tabular}
\end{ruledtabular}
\label{table3}
\end{table*}
\begin{table*}[h]
\caption{Coefficients \protect$\Upsilon^{(ad,na,pv)}_{j,k}$ for the 
         \protect$(\pi\sigma)_1$ contributions.}                          
\begin{ruledtabular}
\begin{tabular}{c|ccc}  
 &&&\\
    & $\Upsilon_{0}(//)(t,u)$ & $\Upsilon_1(//)(t,u)$ & $\Upsilon_2(//)(t,u)$ \\            
&&&\\
\colrule
&&&\\
  & \multicolumn{3}{c}{ 1-pair exchange} \\
&&&\\
\colrule
&&&\\
 $\Omega^{(1),ad}_2$ & $\displaystyle{-\frac{2}{t+u} }$ & 
 $\displaystyle{+\frac{2}{3}\frac{tu-u^2}{(t+u)^2} }$ & --- \\[3mm]      
 $\Omega^{(1),ad}_3$ &  
 $\displaystyle{+2\frac{tu-u^2}{(t+u)^2} }$ & --- & --- \\[3mm]      
&&&\\
 $\Omega^{(1),na}_2$ & $\displaystyle{+\frac{5}{\sqrt{\pi}}
 \frac{1}{(t+u)^{3/2}}\cdot\frac{1}{M} }$ & $\displaystyle{+\frac{1}{3\sqrt{\pi}}
 \frac{t^2-13tu+6u^2}{(t+u)^{5/2}}\cdot\frac{1}{M} }$ & $\displaystyle{
 +\frac{2}{3\sqrt{\pi}}\frac{tu^2(t-u)}{(t+u)^{7/2}}\cdot\frac{1}{M} }$ \\[3mm]
 $\Omega^{(1),na}_3$ & $\displaystyle{ +\frac{1}{\sqrt{\pi}}
 \frac{t^2-7tu+6u^2}{(t+u)^{5/2}}\cdot\frac{1}{M} }$ & $\displaystyle{
 +\frac{2}{\sqrt{\pi}}\frac{tu^2(t-u)}{(t+u)^{7/2}}\cdot\frac{1}{M} }$ & --- \\[3mm]
&&&\\
 $\Omega^{(1),pv}_2$ & $\displaystyle{-\frac{1}{\sqrt{\pi}}
 \frac{1}{(t+u)^{3/2}}\cdot\frac{1}{M} }$ & $\displaystyle{-
 \frac{1}{3\sqrt{\pi}}\frac{t^2-tu}{(t+u)^{5/2}}\cdot\frac{1}{M} }$ & --- \\[3mm]
 $\Omega^{(1),pv}_3$ & $\displaystyle{-
 \frac{1}{\sqrt{\pi}}\frac{t^2-tu}{(t+u)^{5/2}}\cdot\frac{1}{M} }$ & --- & --- \\[3mm]
&&&\\
\colrule
&&&\\
  & \multicolumn{3}{c}{ 2-pair exchange} \\
&&&\\
\colrule
&&&\\
 $\Omega^{(2),ad}_2$ & $\displaystyle{-\frac{1}{\sqrt{\pi}} \frac{1}{(t+u)^{3/2}}}$ & 
 $\displaystyle{-\frac{1}{6\sqrt{\pi}}\frac{(t-u)^2}{(t+u)^{5/2}} }$ & --- \\[3mm]      
 --- & ---- \\[3mm]
 $\Omega^{(2),ad}_3$ & 
 $\displaystyle{-\frac{1}{2\sqrt{\pi}}\frac{(t-u)^2}{(t+u)^{5/2}} }$ &       
 --- & ---- \\[3mm]
\end{tabular}
\end{ruledtabular}
\label{table4}
\end{table*}
\begin{table*}[h]
\caption{Coefficients \protect$\Upsilon^{(ad,na,pv)}_{j,k}$ for the 
         \protect$(\pi\omega)_1$ contributions.}                          
\begin{ruledtabular}
\begin{tabular}{c|ccc}  
 &&&\\
    & $\Upsilon_{0}(//)(t,u)$ & $\Upsilon_1(//)(t,u)$ & $\Upsilon_2(//)(t,u)$ \\            
&&&\\
\colrule
&&&\\
  & \multicolumn{3}{c}{ 1-pair exchange} \\
&&&\\
\colrule
&&&\\
 $\Omega^{(1),ad}_2$ & --- &
 $\displaystyle{+\frac{2}{3}\frac{u}{t+u} }$ & --- \\[3mm]  
 $\Omega^{(1),ad}_3$ & 
 $\displaystyle{+2\frac{u}{t+u} }$ & --- & --- \\[3mm]  
&&&\\
\colrule
&&&\\
  & \multicolumn{3}{c}{ 2-pair exchange} \\
&&&\\
\colrule
&&&\\
 $\Omega^{(2),ad}_2$ & $\displaystyle{-\frac{1}{2\sqrt{\pi}}
 \left[\frac{\delta(t-t_0)}{\sqrt{u}}+\frac{\delta(u-u_0)}{\sqrt{t}}\right] }$ & 
 $\displaystyle{+\frac{1}{6\sqrt{\pi}}\frac{1}{\sqrt{t+u}} }$ & --- \\[3mm]
 $\Omega^{(2),ad}_3$ & $\displaystyle{+\frac{1}{2\sqrt{\pi}}\frac{1}{\sqrt{t+u}} }$ &
 --- & --- \\[3mm]
\end{tabular}
\end{ruledtabular}
\label{table5}
\end{table*}
\begin{table*}[h]
\caption{Coefficients \protect$\Upsilon^{(ad,na,pv)}_{j,k}$ for the 
         \protect$(\pi P)_1$ contributions.
         These coefficients have to be multiplied by a factor $-\delta(u-u_0)/M_N^2$.
 }                          
\begin{ruledtabular}
\begin{tabular}{c|ccc}  
 &&&\\
    & $\Upsilon_{0}(//)(t,u)$ & $\Upsilon_1(//)(t,u)$ & $\Upsilon_2(//)(t,u)$ \\            
&&&\\
\colrule
&&&\\
  & \multicolumn{3}{c}{ 1-pair exchange} \\
&&&\\
\colrule
&&&\\
 $\Omega^{(1),ad}_2$ & $\displaystyle{-\frac{2}{t+u} }$ & 
 $\displaystyle{+\frac{2}{3}\frac{tu-u^2}{(t+u)^2} }$ & --- \\[3mm]      
 $\Omega^{(1),ad}_3$ &  
 $\displaystyle{+2\frac{tu-u^2}{(t+u)^2} }$ & --- & --- \\[3mm]      
&&&\\
 $\Omega^{(1),na}_2$ & $\displaystyle{+\frac{5}{\sqrt{\pi}}
 \frac{1}{(t+u)^{3/2}}\cdot\frac{1}{M} }$ & $\displaystyle{+\frac{1}{3\sqrt{\pi}}
 \frac{t^2-13tu+6u^2}{(t+u)^{5/2}}\cdot\frac{1}{M} }$ & $\displaystyle{
 +\frac{2}{3\sqrt{\pi}}\frac{tu^2(t-u)}{(t+u)^{7/2}}\cdot\frac{1}{M} }$ \\[3mm]
 $\Omega^{(1),na}_3$ & $\displaystyle{ +\frac{1}{\sqrt{\pi}}
 \frac{t^2-7tu+6u^2}{(t+u)^{5/2}}\cdot\frac{1}{M} }$ & $\displaystyle{
 +\frac{2}{\sqrt{\pi}}\frac{tu^2(t-u)}{(t+u)^{7/2}}\cdot\frac{1}{M} }$ & --- \\[3mm]
&&&\\
 $\Omega^{(1),pv}_2$ & $\displaystyle{-\frac{1}{\sqrt{\pi}}
 \frac{1}{(t+u)^{3/2}}\cdot\frac{1}{M} }$ & $\displaystyle{-
 \frac{1}{3\sqrt{\pi}}\frac{t^2-tu}{(t+u)^{5/2}}\cdot\frac{1}{M} }$ & --- \\[3mm]
 $\Omega^{(1),pv}_3$ & $\displaystyle{-
 \frac{1}{\sqrt{\pi}}\frac{t^2-tu}{(t+u)^{5/2}}\cdot\frac{1}{M} }$ & --- & --- \\[3mm]
&&&\\
\colrule
&&&\\
  & \multicolumn{3}{c}{ 2-pair exchange} \\
&&&\\
\colrule
&&&\\
 $\Omega^{(2),ad}_2$ & $\displaystyle{-\frac{1}{\sqrt{\pi}} \frac{1}{(t+u)^{3/2}}}$ & 
 $\displaystyle{-\frac{1}{6\sqrt{\pi}}\frac{(t-u)^2}{(t+u)^{5/2}} }$ & --- \\[3mm]      
 --- & ---- \\[3mm]
 $\Omega^{(2),ad}_3$ & 
 $\displaystyle{-\frac{1}{2\sqrt{\pi}}\frac{(t-u)^2}{(t+u)^{5/2}} }$ &       
 --- & ---- \\[3mm]

\end{tabular}
\end{ruledtabular}
\label{table6}
\end{table*}
\begin{table*}[h]
\caption{Coefficients \protect$\Upsilon^{(ad,na,pv)}_{j,k}$ for the 
         \protect$(\pi \pi)_0('derivative')$ contributions.
 }                          
\begin{ruledtabular}
\begin{tabular}{c|c|c|c|c}  
 &&&\\
    & $\Upsilon_{0}(//)(t,u)$ & $\Upsilon_1(//)(t,u)$ & $\Upsilon_2(//)(t,u)$ 
    & $\Upsilon_{3}(//)(t,u)$ \\            
&&&&\\
\colrule
&&&&\\
 $\Omega^{(1),ad}_1$ & $\displaystyle{-\frac{15}{4}\frac{1}{t+u} }$ & 
 $\displaystyle{-\frac{1}{2}\frac{t^2-8tu+u^2}{(t+u)^3} }$ & $\displaystyle{
 -\frac{t^2u^2}{(t+u)^4} }$ & --- \\[3mm]      
 $\Omega^{(1),na}_1$ & $\displaystyle{-\frac{45}{2\sqrt{\pi}}
 \frac{1}{(t+u)^{5/2}}\cdot\frac{1}{M} }$ & $\displaystyle{+\frac{1}{2\sqrt{\pi}}
 \frac{14t^2-67tu+14u^2}{(t+u)^{7/2}}\cdot\frac{1}{M} }$ & $\displaystyle{
 -\frac{tu}{\sqrt{\pi}}\frac{3t^2-14tu+3u^2}{(t+u)^{9/2}}\cdot\frac{1}{M} }$ 
 & $\displaystyle{ -\frac{1}{\sqrt{\pi}} \frac{t^3u^3}{(t+u)^{1/2}}\cdot
 \frac{1}{M} }$ \\[3mm]
 $\Omega^{(1),na}_4$ & $\displaystyle{ +\frac{9}{\sqrt{\pi}}
 \frac{1}{(t+u)^{3/2}}\cdot\frac{1}{M} }$ & $\displaystyle{
 -\frac{4}{\sqrt{\pi}}\frac{tu}{(t+u)^{5/2}}\cdot\frac{1}{M} }$ & --- 
 & --- \\[3mm]
&&&&\\
 $\Omega^{(1),pv}_1$ & $\displaystyle{-\frac{1}{2\sqrt{\pi}}
 \frac{(m_1^2-m_2^2)^2}{(t+u)^{1/2}}\cdot\frac{1}{M}}$ & $\displaystyle{\frac{1}{2\sqrt{\pi}}
 \frac{m_1^2(3t^2-u^2)-m_2^2(t^2-3u^2)}{(t+u)^{5/2}}
 \cdot\frac{1}{M} }$ & $\displaystyle{ 
 \frac{1}{\sqrt{\pi}}\frac{tu(t^2+2tu+u^2)}{(t+u)^{9/2}} 
 \cdot\frac{1}{M} }$ & --- \\[3mm]
&&&&\\
      & $\displaystyle{+\frac{3}{2\sqrt{\pi}}\frac{m_1^2+m_2^2}{(t+u)^{3/2}}\cdot\frac{1}{M}}$ & $\displaystyle{+\frac{1}{2\sqrt{\pi}}\frac{t^2+2tu+u^2}{(t+u)^{7/2}}    
      \cdot\frac{1}{M} }$ && \\[3mm]
 $\Omega^{(1),pv}_4$ & $\displaystyle{-\frac{2}{\sqrt{\pi}}
 \frac{(m_1^2+m_2^2)}{(t+u)^{1/2}}\cdot\frac{1}{M}}$ & $\displaystyle{-\frac{2}{\sqrt{\pi}}
 \frac{(t^2+tu+u^2)}{(t+u)^{5/2}}
 \cdot\frac{1}{M} }$ & --- & --- \\[3mm]
      & $\displaystyle{-\frac{3}{\sqrt{\pi}}\frac{1}{(t+u)^{3/2}}\cdot\frac{1}{M}}$&&\\[3mm]
&&&&\\
\colrule
&&&&\\
 $\Omega^{(2),ad}_1$ & $\displaystyle{+\frac{1}{\sqrt{\pi}} 
 \frac{1}{(t+u)^{3/2}}}$ & 
 $\displaystyle{-\frac{1}{2\sqrt{\pi}}\frac{tu}{(t+u)^{5/2}} }$ & --- 
 & --- \\[3mm]
\end{tabular}
\end{ruledtabular}
\label{table7p}
\end{table*}

\twocolumngrid
\end{widetext}


\begin{thebibliography}{99}

\bibitem{RN00a}       
 Th.A.\ Rijken, H.\ Polinder, and J.\ Nagata, 
 {\it ESC NN-Potentials in Momentum Space.   
    I. PS-PS Exchange Potentials }, preprint 2000, refered to as paper I.
\bibitem{RKS91}     
 T.A. Rijken, R.A.M.\ Klomp, and J.J. de Swart, 'Soft-Core OBE-Potentials
 in Momentum Space', preprint June 1991, Institute for Theoretical Physics, 
 Nijmegen, The Netherlands.
\bibitem{RS96a}      
 Th.A.\ Rijken and V.G.J.\ Stoks, Phys.\ Rev.\ {\bf 54}, 2851 (1996).
\bibitem{RS96b}      
 Th.A.\ Rijken and V.G.J.\ Stoks, Phys.\ Rev.\ {\bf 54}, 2869 (1996).
\bibitem{NRS78}      
 M.M. Nagels, T.A. Rijken, and J.J. de Swart,
 Phys.\ Rev.\ D{\bf 17}, 768 (1978).
\bibitem{MRS89}      
 P.M.M. Maessen, T.A. Rijken, and J.J. de Swart,
 Phys.\ Rev.\ D{\bf 40}, 2226 (1989).
\bibitem{Rij85}
 T.A. Rijken, Ann.\ Phys.\ (NY) {\bf 164}, 1 and 23 (1985).
\bibitem{Rij93}      
 Th.A.\ Rijken, {\it Proceedings of the XIVth European Conference on   
 Few-Body Problems in Physics}, Amsterdam 1993, eds. B.\ Bakker and 
 R.\ von Dantzig, Few-Body Systems, Suppl. {\bf 7}, 1 (1994).
\bibitem{SR97}      
 V.G.J.\ Stoks and Th.A. Rijken, Nucl.\ Phys.\ {\bf A613}, (1997) 311.
\bibitem{Rij99}      
 Th.A.\ Rijken, {\it Proceedings of the 1st Asian-Pacific Conference on   
 Few-Body Problems in Physics}, Tokyo 1999, to be published.           
\bibitem{Rij91}      
 Th.A.\ Rijken, Ann.\ Phys.\ (N.Y.) {\bf 208}, 253 (1991).              
\bibitem{Rij00}      
 Th.A.\ Rijken, {\it Derivative Scalar-Pair Exchange Nucleon-nucleon    
 Potentia}, report University ogf Nijmegen, 2001 (unpublished).
\bibitem{Ko94}          
 P.\ Ko and S.\ Rudaz, Phys.\ Rev.\ {\bf D 50} (1994) 6877.         
\bibitem{Ordonez92}      
 C. Ord\'{o}\~{n}ez and U. van Kolck, Phys. Lett. B {\bf 291} (1992)
 459; 
 C. Ord\'{o}\~{n}ez, L. Ray, and U. van Kolck, Phys. Rev.\ Lett.\  
\bibitem{Sto93}          
 V.G.J.\ Stoks, R.A.M.\ Klomp, M.C.M.\ Rentmeester, and J.J.\ de Swart,
 Phys.\ Rev.\ C{\bf 48} (1993) 792.
\bibitem{Bugg92}         
 D.V.\ Bugg and R.A.\ Bryan, Nucl.\ Phys.\ {\bf A540} (1992) 449.
\bibitem{Klo93}          
 R.A.M.\ Klomp, private communication (unpublisjed).    
\bibitem{SKTS94}          
 V.G.J.\ Stoks, R.A.M.\ Klomp, C.P.F.\ Terheggen, and J.J.\ de Swart,
 Phys.\ Rev.\ C{\bf 49} (1994) 2950.
\bibitem{Arn65}          
 R.C.\ Arnold, Phys.\ Rev.\ Lett.\ {\bf 14} (1965) 657;
 C.\ Schmid, Lettere al Nuov.\ Cim.\ {\bf 1} (1969) 165;
 H.J.\ Lipkin, Nucl.\ Phys.\ {\bf B9} (1969) 349.

\end{thebibliography}
\end{document}